# Analysis and applications of the upwind conservation element and solution element scheme for compressible flow simulations

Yazhong Jiang,[1] Lisong Shi,[2,a)] and Chih-Yung Wen[2,a)]

[1] *Department of Mechanics and Engineering Sciences, School of Physics and Mechanics, Wuhan University of Technology, Wuhan, 430070, China*

[2] *Department of Aeronautical and Aviation Engineering, Faculty of Engineering, The Hong Kong Polytechnic University, Hong Kong, 999077, China*

The upwind conservation element and solution element (CESE) scheme is an alternative discontinuity-capturing numerical approach to solving hyperbolic conservation laws. To evaluate the numerical properties of this spatiotemporal coupled scheme, a formal analysis is conducted on the upwind CESE discretization applied to the linear advection problem. The modified equation and the effective modified wavenumber are derived, which theoretically confirm the order of accuracy and reveal the dissipation and dispersion properties of this scheme. Several examples are considered to demonstrate the capabilities of the upwind CESE scheme for simulating compressible flows, including shock–vortex and shock–bubble interactions. The results of the present scheme agree well with exact solutions, results of other numerical methods, and experimental data. This demonstrates the high resolution of the scheme in capturing shock waves, material interfaces, and small-scale flow structures.

## I. INTRODUCTION

The space–time conservation element and solution element (CESE) method,[1,2] initially proposed by Chang and To,[1] is a special finite-volume type of numerical framework for solving equations of conservation laws. As a fully discrete and explicit time-marching method, CESE provides accurate simulations of time-varying physical processes involving the evolution of waves and discontinuities. Examples include the propagation and interaction of shock waves,[3] aeroacoustics,[4] detonations,[5] interfacial instabilities,[6] compressible multi-fluid flows,[7] thermochemical nonequilibrium reentry flows,[8] and phenomena in magnetohydrodynamics (MHD).[9]

In the CESE method, the unknown physical fields and their spatial derivatives are regarded as independent time-marching variables, which are stored at mesh points and updated in each time step. Other features of this method include a unified treatment of space and time, a staggered time-marching strategy using a staggered mesh, and a single-stage time integration based on the Cauchy–Kowalewski procedure. By combining these techniques, the CESE method maintains the strict preservation of space–time coupling and conservativeness, while achieving low dissipation and high resolution with a very compact stencil. Recent reviews of the CESE method provide more details of this method and its applications.[10,11]

[a)] Authors to whom correspondence should be addressed: ls.mark.shi@gmail.com and cywen@polyu.edu.hk

In the past decades, significant advancements have been made in the development of the CESE method. First, the central CESE schemes[2,12–16] have proved to be successful examples of shock-capturing central schemes. Second, the multidimensional CESE schemes[17–22] have demonstrated great flexibility and adaptivity in handling complex geometries and arbitrary meshes. Last but not least, the high-order CESE schemes[16,23–26] have garnered increasing interest as they are capable of achieving arbitrary uniform high order of accuracy in both space and time, with a compact stencil and single-stage time integration.

The upwind CESE method, proposed by Shen, Wen, and Zhang,[27] represents a recent advancement of the original CESE framework. In contrast to Chang's central CESE schemes,[2,13] the upwind CESE schemes[27–30] successfully incorporate upwind numerical flux techniques, such as limiters and approximate Riemann solvers, which are widely used in compressible flow simulations. Consequently, when simulating challenging problems such as detonations,[31] multiphase flows,[32] and MHD flows,[30,33] the upwind CESE schemes exhibit improved accuracy and robustness in capturing discontinuities, particularly strong contact discontinuities (e.g., material interfaces with large density ratios and very different thermodynamic properties[7]).

Since the introduction of the upwind CESE method, many numerical tests have been conducted to demonstrate its capabilities. However, because of the distinctive formulation of this scheme compared to widely used schemes in computational fluid dynamics, the performances observed in specific numerical cases are not easily interpretable. Thus, the intrinsic characteristics of the upwind CESE method have not been well understood. It is imperative and intriguing to perform a formal analysis on the algebraic equations produced by the upwind CESE discretization. This analysis is expected to explain the numerical behavior of the upwind CESE scheme to some extent and highlight its strengths. Moreover, it will offer theoretical and quantitative evidence for an objective evaluation of this method, particularly in comparison to other numerical methods.

To the best of the authors' knowledge, there is still insufficient analysis on the properties of the upwind CESE scheme. Shen et al.[27] proved the Courant-number-insensitive nature of the upwind CESE scheme. Jiang et al.[10] conducted a von Neumann stability analysis of this scheme. However, the properties related to its numerical accuracy have not been theoretically quantified. In this work, the conventional procedures for deriving the modified equation and analyzing spectral properties are extended to account for the unique discrete form of the upwind CESE scheme. The analysis results for this scheme include the order of convergence rate, the truncation error terms, and the numerical dissipation/dispersion properties.

The remainder of this paper is organized as follows. The upwind CESE scheme is described in Sec. II, followed by an analysis of the scheme applied to the linear advection equation in Sec. III. To further validate the upwind CESE scheme and demonstrate its applications to compressible flows, selected examples are presented in Sec. IV, focusing on the dynamics and interactions of shock waves, contact discontinuities, and vortices. Finally, concluding remarks are provided in Sec. V.

## II. CONSTRUCTION OF UPWIND CESE SCHEMES

The upwind CESE method was proposed by Shen et al.[27] and its construction has been detailed in Refs. 10,11, 27–30. This section provides a brief review describing the upwind CESE schemes applied for solving different governing equations.

### A. Upwind CESE scheme for 1D scalar conservation law

First, the upwind CESE scheme is described through its application to a 1D scalar conservation law given by

$$\frac{\partial u}{\partial t}+\frac{\partial f(u)}{\partial x}=0, \tag{1}$$

where $u$ is the conserved variable, and $f(u)$ is the flux function. Note that throughout this paper, partial derivatives are also expressed in subscript form, e.g., $u_x$ represents $\partial u/\partial x$.

#### 1. *Discretization and related notations*

The $x$–$t$ plane in Fig. 1 is discretized using a mesh of solid lines with a uniform cell size $\Delta x$ and a time step size $\Delta t$. At the same time, the CESE method utilizes a staggered space–time mesh represented by the dashed lines in Fig. 1. In a second-order CESE scheme, discrete values of $u$ and $u_x$ are calculated and stored at all mesh points, including points $(x_j, t_n)$ and $(x_{j-1/2}, t_{n-1/2})$ indexed by integers and half-integers, respectively. The time-marching algorithm progresses from $t_{n-1}$ to $t_n$ through two half steps. During the first half step from $t_{n-1}$ to $t_{n-1/2}$, the unknowns $u_{j-1/2}^{n-1/2}$ and $(u_x)_{j-1/2}^{n-1/2}$ are derived using the information stored at points $(j-1, n-1)$ and $(j, n-1)$. In the second half step from $t_{n-1/2}$ to $t_n$, the unknowns $u_j^n$ and $(u_x)_j^n$ are derived using the information stored at points $(j-1/2, n-1/2)$ and $(j+1/2, n-1/2)$. The arrows in Fig. 1 illustrate the flow of information in a second-order CESE scheme.

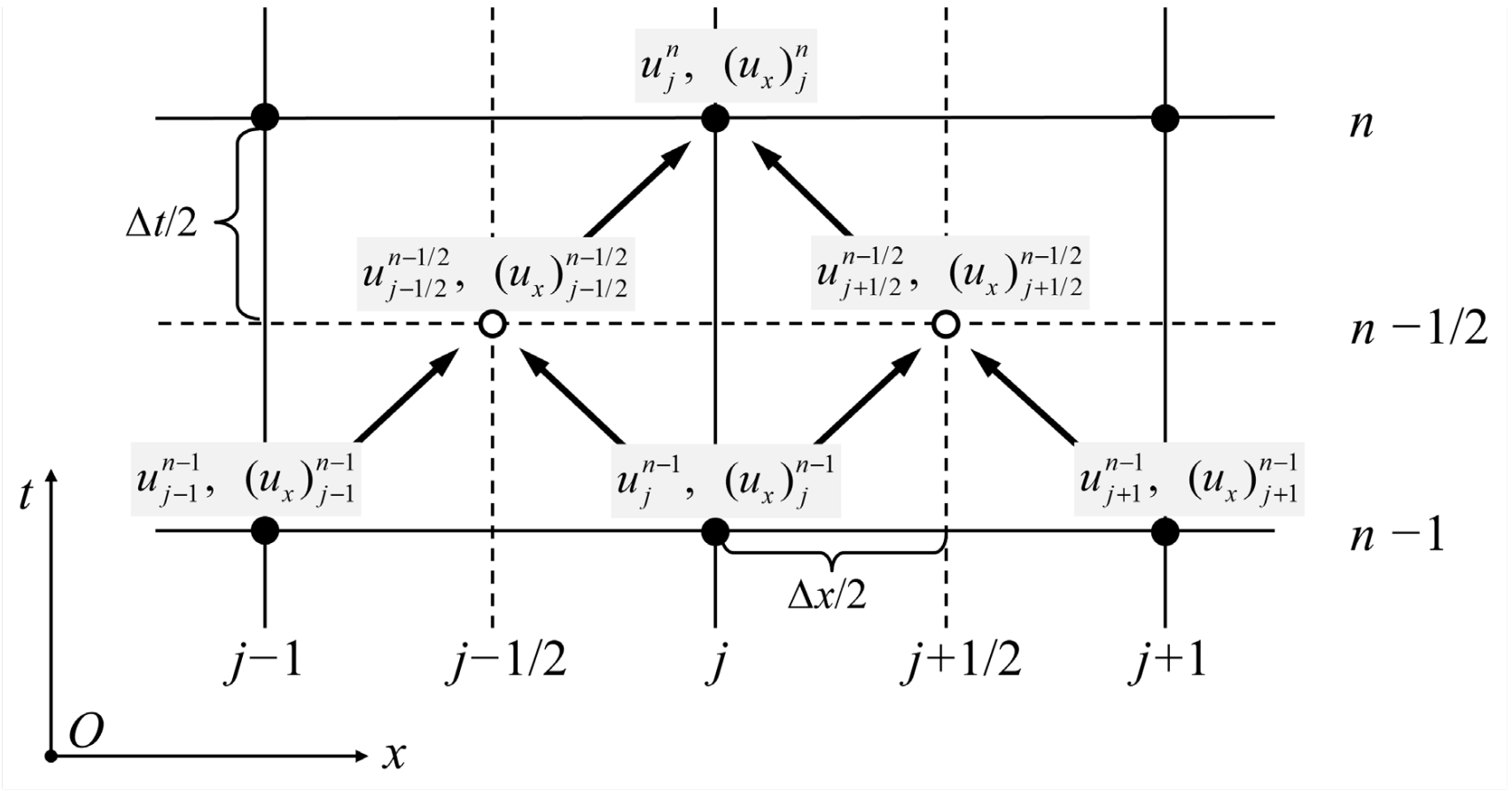

FIG. 1. Computational mesh, time-marching variables, and the flow of information in the CESE method.

To conveniently explain the ideas behind this numerical scheme, two important concepts are introduced: the conservation element (CE) and the solution element (SE). As shown in Fig. 2, an SE represents the elementary space–time domain in which the solution is approximated by a local polynomial. As the CESE method stores and evolves the conserved variables and their derivatives, the interior of $(\mathrm{SE})_j^n$ can be formed through a Taylor expansion. In the case of the second-order scheme discussed here, $u$ and $f$ inside $(\mathrm{SE})_j^n$ are constructed by first-order Taylor expansions at ($x_j$, $t_n$) as follows:

$$u(x,t)=u_j^n+(u_x)_j^n(x-x_j)+(u_t)_j^n(t-t_n), \qquad x_{j-1/2}<x<x_{j+1/2},\ \ t_n\le t<t_{n+1/2}, \tag{2}$$

$$f(x,t)=f_j^n+(f_x)_j^n(x-x_j)+(f_t)_j^n(t-t_n), \qquad x_{j-1/2}<x<x_{j+1/2},\ \ t_n\le t<t_{n+1/2}. \tag{3}$$

All these non-overlapping SEs cover the $x$–$t$ plane, and the SEs are staggered for every two successive half steps. Hence, the entire solution can be constructed by assembling the solutions within the SEs (Eq. (2)). Discontinuities in the solution are allowed at the interfaces (e.g., the line segment CD in Fig. 2) between two adjacent SEs.

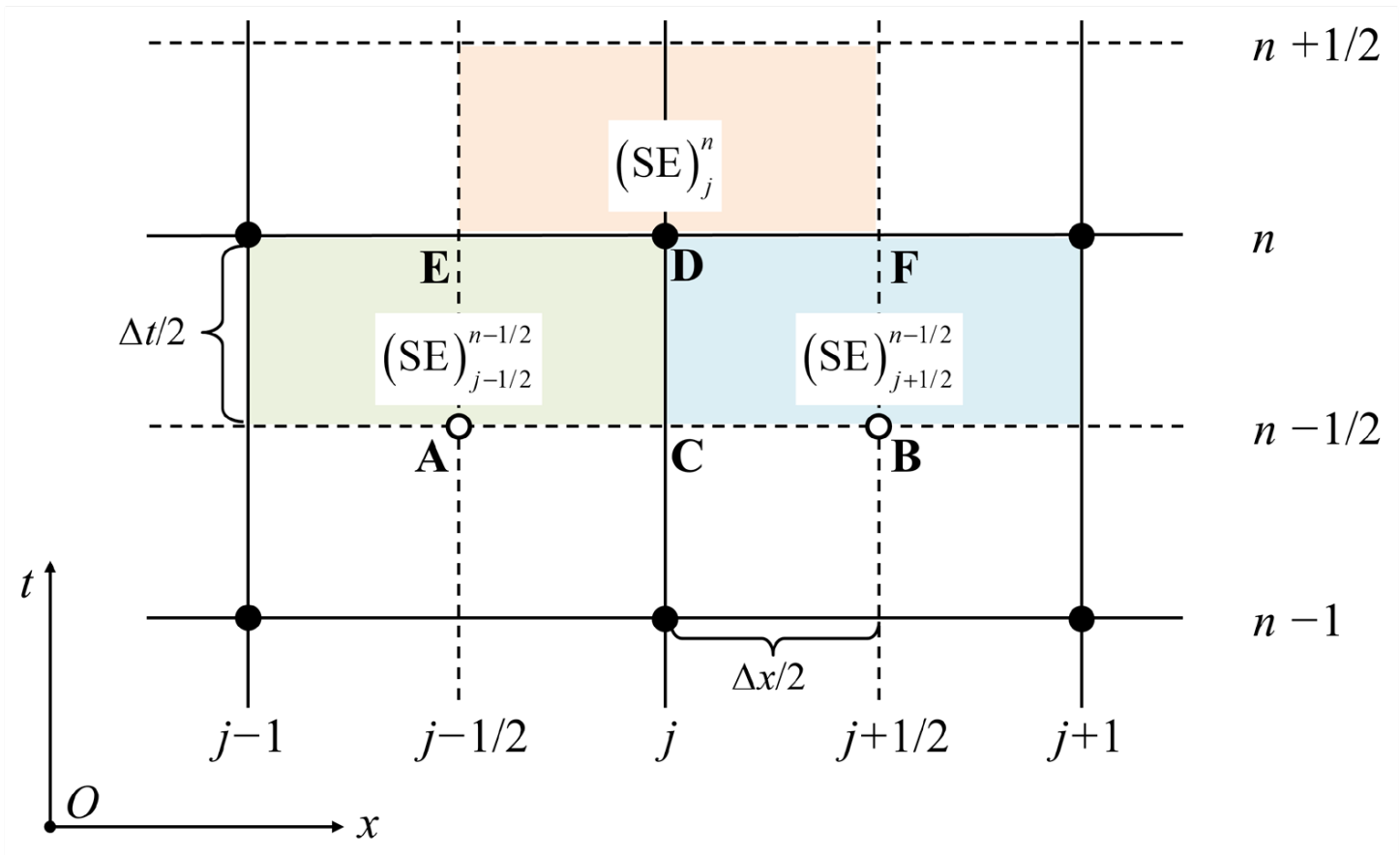

FIG. 2. Definition of the solution element (SE) in the upwind CESE scheme.

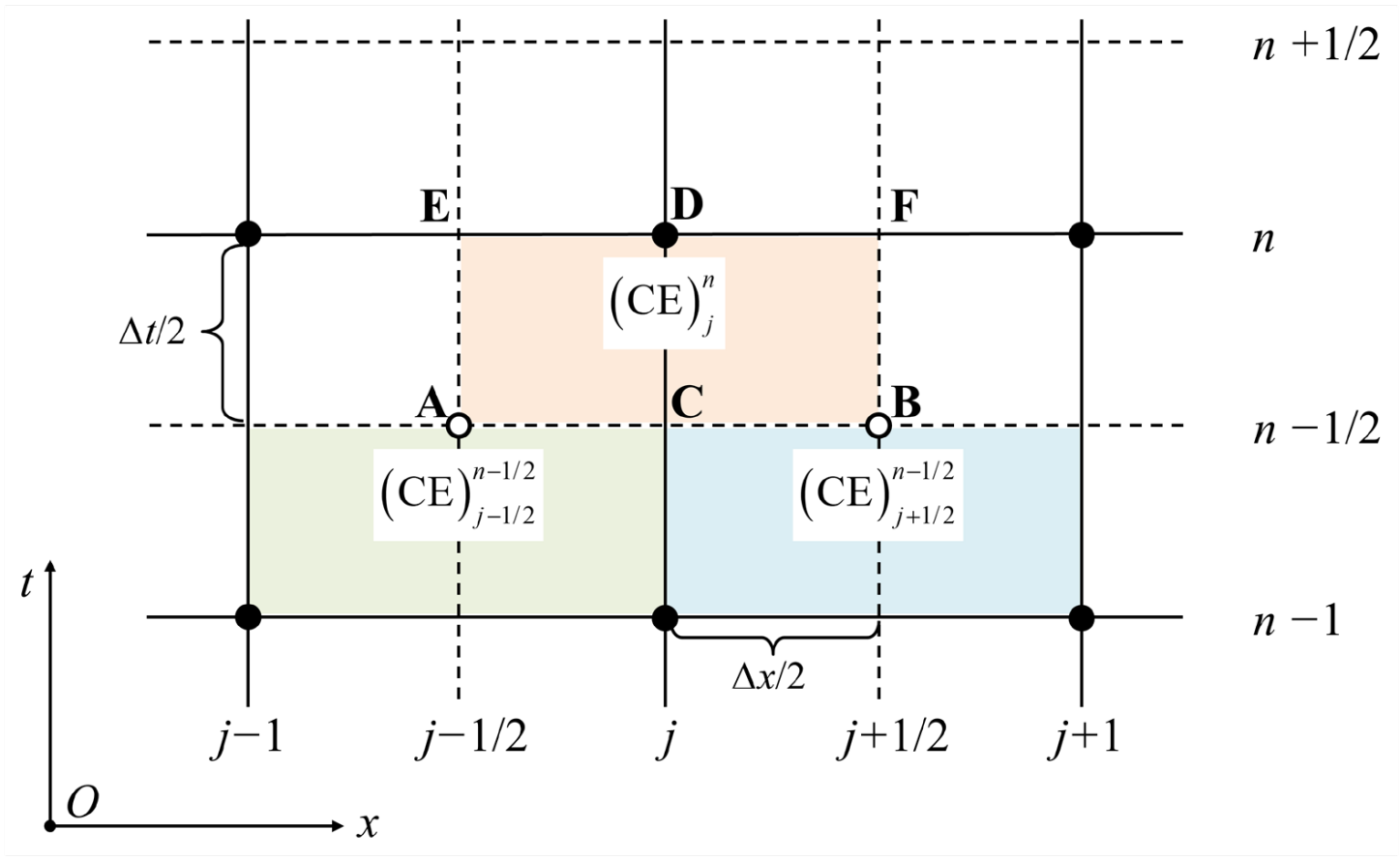

FIG. 3. Definition of the conservation element (CE) in the upwind CESE scheme.

As shown in Fig. 3, a CE represents the elementary space–time domain where the conservation equation is integrated. Thus, the space–time integral form of the conservation law is numerically implemented over each CE. A CE is assigned to every solution point. For point ( *j*, *n*), its corresponding CE is denoted by $(\mathrm{CE})_j^n$, illustrated as the rectangle ABFE in Fig. 3. These CEs collectively cover the *x*–*t* plane in a staggered manner, while ensuring they do not overlap.

A comparison between Fig. 2 and Fig. 3 reveals the dislocation between the CEs and SEs. A CE never coincides with an SE, and the staggered arrangement of CEs and SEs is specially designed. Consequently, the boundaries of $(\mathrm{CE})_j^n$ belong to three different SEs: AC and AE belong to $(\mathrm{SE})_{j-1/2}^{n-1/2}$, BC and BF belong to $(\mathrm{SE})_{j+1/2}^{n-1/2}$, and DE and DF belong to $(\mathrm{SE})_j^n$. The line segment CD, which represents the interface between $(\mathrm{SE})_{j-1/2}^{n-1/2}$ and $(\mathrm{SE})_{j+1/2}^{n-1/2}$, divides $(\mathrm{CE})_j^n$ into two sub-CEs. As shown in Fig. 4, the resulting sub-CEs are denoted by $(\mathrm{CE}^-)_j^n$ and $(\mathrm{CE}^+)_j^n$.

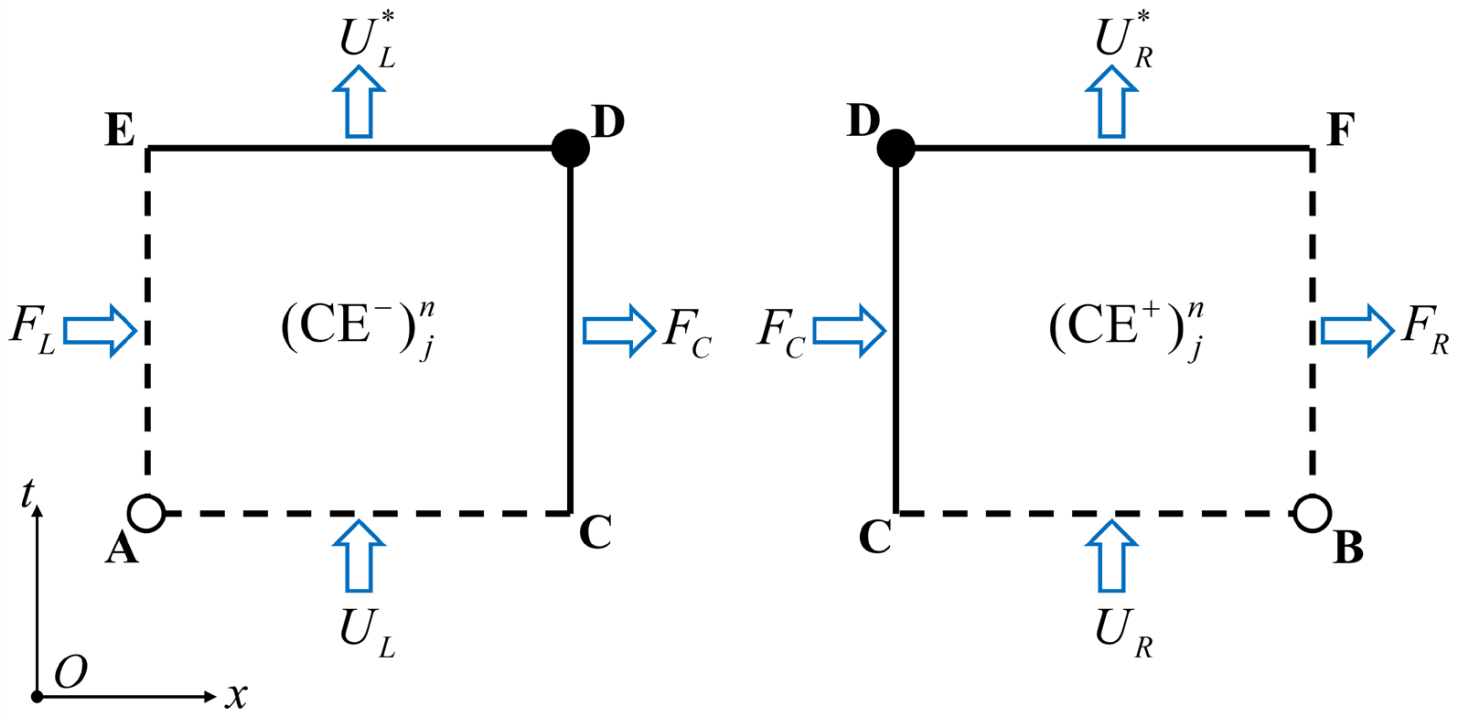

FIG. 4. Definition of sub-CEs (CE$^-$ and CE$^+$), and the notation for the average value of $u(x, t)$ or $f(x, t)$ along each line segment.

## 2. *Equations for each conservation element*

Consider a half step marching from time level $n-1/2$ to time level $n$. At point ( *j*, *n*), two unknowns, $u_j^n$ and $(u_x)_j^n$, must be solved for. Thus, two equations must be established. First, Eq. (1) is integrated over the sub-CEs, $(\mathrm{CE}^-)_j^n$ and $(\mathrm{CE}^+)_j^n$, respectively. By applying Gauss's divergence theorem, these integral conservation equations can be expressed as follows:

$$U_L^* \frac{\Delta x}{2} = U_L \frac{\Delta x}{2} + (F_L - F_C)\frac{\Delta t}{2}, \tag{4}$$

$$U_R^* \frac{\Delta x}{2} = U_R \frac{\Delta x}{2} + (F_C - F_R)\frac{\Delta t}{2}, \tag{5}$$

where the meanings of $U_L^*$, $U_R^*$, $U_L$, $U_R$, $F_L$, $F_R$, and $F_C$ are depicted in Fig. 4.

The average values of $u(x, t)$ along line segments DE, DF, AC, and BC are denoted by $U_L^*$, $U_R^*$, $U_L$, and $U_R$, respectively. Furthermore, $F_L$, $F_R$, and $F_C$ represent the average values of $f(x, t)$ along line segments AE, BF, and CD, respectively.

## 3. Calculation of $U_L$, $U_R$, $F_L$, and $F_R$

The concept of the SE is used here to help calculate the right-hand sides of Eqs. (4) and (5). Recall that AC and AE belong to $(\mathrm{SE})_{j-1/2}^{n-1/2}$. By incorporating the assumed Taylor expansions of $u$ and $f$ in $(\mathrm{SE})_{j-1/2}^{n-1/2}$ into the definitions of $U_L$ and $F_L$, the expressions of $U_L$ and $F_L$ can be derived based on the information at point ($j-1/2$, $n-1/2$) as

$$U_L = u_{j-1/2}^{n-1/2} + \frac{\Delta x}{4}(u_x)_{j-1/2}^{n-1/2}, \tag{6}$$

$$F_L = f_{j-1/2}^{n-1/2} + \frac{\Delta t}{4}(f_t)_{j-1/2}^{n-1/2}. \tag{7}$$

Similarly, as BC and BF lie in $(\mathrm{SE})_{j+1/2}^{n-1/2}$, the Taylor expansions at point ($j+1/2$, $n-1/2$) lead to

$$U_R = u_{j+1/2}^{n-1/2} - \frac{\Delta x}{4}(u_x)_{j+1/2}^{n-1/2}, \tag{8}$$

$$F_R = f_{j+1/2}^{n-1/2} + \frac{\Delta t}{4}(f_t)_{j+1/2}^{n-1/2}. \tag{9}$$

In Eqs. (6)–(9), $u$ and $u_x$ at time level $n-1/2$ are known. The CESE method also requires calculating $f$, $f_x$, $u_t$, and $f_t$ based on the values of $u$ and $u_x$ at each solution point. Utilizing the flux function $f(u)$ prescribed in the governing equation yields

$$f_{j\pm1/2}^{n-1/2} = f(u_{j\pm1/2}^{n-1/2}), \tag{10}$$

$$(f_x)_{j\pm1/2}^{n-1/2} = \left[(\partial f/\partial u)u_x\right]_{j\pm1/2}^{n-1/2}. \tag{11}$$

Next, the time derivative can be obtained by using the governing equation itself. Therefore,

$$(u_t)_{j\pm1/2}^{n-1/2} = -(f_x)_{j\pm1/2}^{n-1/2}. \tag{12}$$

Then, $f_t$ at the same point can be derived as

$$(f_t)_{j\pm1/2}^{n-1/2} = \left[(\partial f/\partial u)u_t\right]_{j\pm1/2}^{n-1/2}. \tag{13}$$

Thus far, $U_L$, $U_R$, $F_L$, and $F_R$ can be explicitly evaluated by using Eqs. (6)–(9).

## 4. Calculation of $F_C$

To proceed with the time-marching scheme from time level $n-1/2$ to time level $n$, the unknown flux $F_C$ in Eqs. (4) and (5) must be addressed. As shown in Fig. 4, $F_C$ represents the average flux through the "diaphragm" CD inside the conservation element $(\mathrm{CE})_j^n$. Note that CD is also the interface between two different solution elements $(\mathrm{SE})_{j-1/2}^{n-1/2}$ and $(\mathrm{SE})_{j+1/2}^{n-1/2}$, making it appropriate to calculate $F_C$ by approximately solving a local Riemann problem at the midpoint of line segment CD (referred to as point $(j,\ n-1/4)$). The discontinuous initial data on the left and right sides of the "diaphragm" CD can be constructed as follows. By using the Taylor expansion in $(\mathrm{SE})_{j-1/2}^{n-1/2}$, the left state at point $(j,\ n-1/4)$ is

$$\left(u_L\right)_j^{n-1/4} = U_L + \frac{\Delta x}{4} u_x^L + \frac{\Delta t}{4} u_t^L , \tag{14}$$

and the Taylor expansion in $(\mathrm{SE})_{j+1/2}^{n-1/2}$ provides the right state at point $(j,\ n-1/4)$:

$$\left(u_R\right)_j^{n-1/4} = U_R - \frac{\Delta x}{4} u_x^R + \frac{\Delta t}{4} u_t^R , \tag{15}$$

where $U_L$ and $U_R$ are defined by Eqs. (6) and (8), respectively. Moreover, $u_x^L$, $u_x^R$, $u_t^L$, and $u_t^R$ in Eqs. (14) and (15) represent the limited derivatives. Using limiters for these derivatives is crucial to suppress the occurrence of spurious oscillations in the presence of discontinuities. The limited slopes can be expressed as

$$u_x^L = (u_x)_{j-1/2}^{n-1/2} \cdot \psi_L, \qquad u_x^R = (u_x)_{j+1/2}^{n-1/2} \cdot \psi_R , \tag{16}$$

where $\Psi$ denotes any appropriate limiter. In this paper, the weighted biased averaging procedure (WBAP) limiter[34] is adopted. Specifically, the limiting procedure is implemented as follows:

$$\psi_L = W(1,\ \theta_1^L,\ \theta_2^L), \qquad \theta_1^L = \frac{(U_R - U_L)/(\Delta x/2)}{(u_x)_{j-1/2}^{n-1/2}}, \qquad \theta_2^L = \frac{(u_x)_{j+1/2}^{n-1/2}}{(u_x)_{j-1/2}^{n-1/2}} , \tag{17}$$

$$\psi_R = W(1,\ \theta_1^R,\ \theta_2^R), \qquad \theta_1^R = \frac{(U_R - U_L)/(\Delta x/2)}{(u_x)_{j+1/2}^{n-1/2}}, \qquad \theta_2^R = \frac{(u_x)_{j-1/2}^{n-1/2}}{(u_x)_{j+1/2}^{n-1/2}} , \tag{18}$$

with the limiter function $W$

$$W(1,\ \theta_1,\ \theta_2) = \begin{cases} \dfrac{5 + 1/\theta_1 + 1/\theta_2}{5 + 1/\theta_1^2 + 1/\theta_2^2}, & \text{if } \theta_1 > 0 \text{ and } \theta_2 > 0 \\ 0, & \text{otherwise} \end{cases} . \tag{19}$$

Based on the spatial derivatives $u_x^L$ and $u_x^R$, the temporal derivatives $u_t^L$ and $u_t^R$ can be obtained using the techniques shown in Eqs. (11) and (12), and this conversion yields

$$u_t^L = -u_x^L \left.\frac{\partial f}{\partial u}\right|_{u=U_L}, \qquad u_t^R = -u_x^R \left.\frac{\partial f}{\partial u}\right|_{u=U_R}. \tag{20}$$

The results of Eqs. (14) and (15) then provide the left and right data for the local Riemann problem. By employing a suitable upwind flux solver $\hat{F}$, the inner flux $F_C$ for $(\mathrm{CE})_j^n$ can be evaluated as

$$F_C = \hat{F}\left((u_L)_j^{n-1/4}, (u_R)_j^{n-1/4}\right). \tag{21}$$

### 5. Time marching of u and $u_x$

Upon completing the calculations of $U_L$, $U_R$, $F_L$, $F_R$, and $F_C$, the integral conservation equations (Eqs. (4) and (5)) readily provide the values of $U_L^*$ and $U_R^*$ at the new time level $t_n$. As $U_L^*$ and $U_R^*$ represent the average values of $u$ along line segments that belong to $(\mathrm{SE})_j^n$, the unknowns $u_j^n$ and $(u_x)_j^n$ can be related to $U_L^*$ and $U_R^*$ through a Taylor expansion inside $(\mathrm{SE})_j^n$:

$$U_L^* = u_j^n - \frac{\Delta x}{4}(u_x)_j^n, \tag{22}$$

$$U_R^* = u_j^n + \frac{\Delta x}{4}(u_x)_j^n. \tag{23}$$

By combining the two equations above, the values of $u_j^n$ and $(u_x)_j^n$ can eventually be determined.

Some remarks are provided below to shed light on the features of the upwind CESE scheme. Substituting Eqs. (4) and (5) into Eqs. (22) and (23) results in

$$u_j^n = \frac{1}{2}(U_L + U_R) + \frac{\Delta t}{2\Delta x}(F_L - F_R) \tag{24}$$

$$\frac{\Delta x}{4}(u_x)_j^n = \frac{1}{2}(U_R - U_L) + \frac{\Delta t}{2\Delta x}(2F_C - F_L - F_R) \tag{25}$$

Among the fluxes $F_L$, $F_R$, and $F_C$, only the inner flux $F_C$ is dependent on the upwind procedure. Furthermore, $F_C$ is precisely cancelled out when deriving Eq. (24), so the upwind procedure does not directly affect the evolution of the cell average of $u$. The role of the upwind procedure is to redistribute $u$ within the cell by determining the slope $u_x$ using Eq. (25).

## B. Upwind CESE scheme for 1D Euler equation

The 1D unsteady compressible Euler equation for a calorically perfect gas can be written as

$$\frac{\partial \boldsymbol{U}}{\partial t}+\frac{\partial \boldsymbol{F}}{\partial x}=\mathbf{0}, \tag{26}$$

where $\boldsymbol{U}$ is the vector of conserved variables $\boldsymbol{U} = [\,\rho, \rho u, E\,]^{\mathrm{T}}$, and $\boldsymbol{F}$ is the inviscid flux vector $\boldsymbol{F}(\boldsymbol{U}) = [\,\rho u, \rho u^2 + p, (E + p)u\,]^{\mathrm{T}}$. The primitive variables $\rho$, $u$, $p$, and $E$ represent the density, velocity, pressure, and total energy per unit volume of the fluid, respectively. The expression for the total energy is $E = p/(\gamma - 1) + \rho u^2/2$, with a constant ratio of specific heat $\gamma$.

The method outlined in Sec. II A for scalar conservation laws can also be applied to Eq. (26). Extending the scheme to a vectorial conservation equation is relatively straightforward, with the exception of the treatment of the local Riemann problem at the interface between sub-CEs: $(\mathrm{CE}^-)_j^n$ and $(\mathrm{CE}^+)_j^n$. The complexity of the Euler equation (Eq. (26)) necessitates the use of approximate Riemann solvers[35,36] for calculating the flux $\boldsymbol{F}_C$, which can be expressed as

$$\boldsymbol{F}_C=\hat{\boldsymbol{F}}\left((\boldsymbol{U}_L)_j^{n-1/4},(\boldsymbol{U}_R)_j^{n-1/4}\right), \tag{27}$$

where $\hat{\boldsymbol{F}}$ represents any upwind flux function suitable for the compressible Euler equation. The discontinuous initial data $(\boldsymbol{U}_L)_j^{n-1/4}$ and $(\boldsymbol{U}_R)_j^{n-1/4}$ are determined in a similar manner as described in Eqs. (14)–(20), with certain limitations imposed on the spatial and temporal slopes of $\boldsymbol{U}(x, t)$. In this work, the contact-restoring Harten, Lax, and van Leer (HLLC) Riemann solver[37] and the WBAP limiter[34] are employed in the upwind procedure for $\boldsymbol{F}_C$.

## C. Upwind CESE scheme for multi-dimensional and multi-fluid problems

The upwind CESE scheme for 1D problems has been successfully extended to multi-dimensional compressible Euler and Navier–Stokes equations.[20,28] The mesh can be either structured or unstructured. Both triangular and quadrilateral cells can be used. To solve a multi-dimensional problem, the independent time-marching variables at each solution point include the conserved variables and their derivatives with respect to all spatial coordinates. For example, a 2D scheme involves storing and evolving $\boldsymbol{U}$, $\boldsymbol{U}_x$, and $\boldsymbol{U}_y$. In the time-marching algorithm, the fluxes through the diaphragms inside each CE can be calculated using an upwind procedure. Here, the rotated HLLC Riemann solver[38–40] is implemented for improved accuracy and robustness.

Compressible multi-fluid flows can be described by a volume-fraction-based five-equation model[41] with the stiffened-gas equation of state.[42] The upwind CESE scheme extended to this model[7] can capture discontinuities with high resolution. It has also been shown to maintain mass conservation for each fluid component and ensure the positivity of volume fractions.

## III. NUMERICAL PROPERTIES

For simplicity and without loss of generality, the one-dimensional linear scalar advection equation

$$u_t + au_x = 0, \quad (a > 0) \tag{28}$$

is considered in this section. In the equation above, $u_t$ and $u_x$ denote the temporal and spatial derivatives of $u(x, t)$, respectively. The wave speed $a$ is assumed to be a positive constant.

The upwind CESE scheme for this equation is simply a special case of the numerical scheme described in Sec. II A, where $f(u)$ in Eq. (1) is set to be $au$. Because of the simplicity of the linear advection, the inner flux $F_C$ has an exact expression:

$$F_C = a(u_L)_j^{n-1/4}. \tag{29}$$

If the slope limiter is not used, the time-marching scheme of $u$ and $u_x$ from time level $n-1/2$ to time level $n$, which is the combination of Eqs. (24) and (25), can be expressed explicitly. By introducing a vector of time-marching variables

$$\boldsymbol{q}_j^n = \begin{bmatrix} u_j^n \\ \dfrac{\Delta x}{4}(u_x)_j^n \end{bmatrix}, \tag{30}$$

the upwind CESE scheme for Eq. (28) can be written in the following form:

$$\boldsymbol{q}_j^n = \boldsymbol{Q}_L \boldsymbol{q}_{j-1/2}^{n-1/2} + \boldsymbol{Q}_R \boldsymbol{q}_{j+1/2}^{n-1/2}, \tag{31}$$

where $j$ and $n$ can be either integers or half-integers. The coefficient matrices in Eq. (31) are

$$\boldsymbol{Q}_L = \frac{1}{2}\begin{bmatrix} 1+\nu & 1-\nu^2 \\ -1+\nu & -1+4\nu-\nu^2 \end{bmatrix}, \qquad \boldsymbol{Q}_R = \frac{1}{2}\begin{bmatrix} 1-\nu & -1+\nu^2 \\ 1-\nu & -1+\nu^2 \end{bmatrix}, \tag{32}$$

where $\nu$ is the Courant–Friedrichs–Lewy (CFL) number defined as

$$\nu \equiv a\frac{\Delta t}{\Delta x}. \tag{33}$$

The matrix form, as shown in Eqs. (30)–(32), characterizes the upwind CESE scheme and acts as the starting point for analyzing its properties in the rest of this section.

Standard approaches for analyzing the stability, the modified equation, and the numerical dissipation and dispersion of a finite-difference scheme for the advection equation are well-established. These analytical procedures can also be extended to the upwind CESE scheme described in this section. It is worth noting that in the CESE scheme, both $u$ and $u_x$ must be treated as two independent degrees of freedom at every solution point.

### A. Stability

A von Neumann stability analysis for the upwind CESE scheme was presented in Ref. 10. This stability analysis is based on the matrix form of Eq. (31) with coefficient matrices shown in Eq. (32). The conclusion was that the scheme is numerically stable when $0 \le v \le 1$. While the specifics of this analysis will not be reiterated here, it is useful to examine the two limiting cases of $v = 0$ and $v = 1$.

(i) $v = 0$. Note that Eq. (31) can also be applied to the half step from $t_{n-1}$ to $t_{n-1/2}$. Then, by combining two successive half steps, the time-marching formula for the complete time step from $t_{n-1}$ to $t_n$ can be obtained as

$$\boldsymbol{q}_j^n = (\boldsymbol{Q}_L)^2 \boldsymbol{q}_{j-1}^{n-1} + (\boldsymbol{Q}_L \boldsymbol{Q}_R + \boldsymbol{Q}_R \boldsymbol{Q}_L) \boldsymbol{q}_j^{n-1} + (\boldsymbol{Q}_R)^2 \boldsymbol{q}_{j+1}^{n-1} . \tag{34}$$

Substituting Eq. (32) with $v = 0$ into Eq. (34) yields $\boldsymbol{q}_j^n = \boldsymbol{q}_j^{n-1}$, which agrees with the zero time advancement implied by $v = 0$.

(ii) $v = 1$. In this case, Eq. (32) shows that $\boldsymbol{Q}_L = \boldsymbol{I}$ and $\boldsymbol{Q}_R = \boldsymbol{O}$, thus simplifying Eq. (34) to $\boldsymbol{q}_j^n = \boldsymbol{q}_{j-1}^{n-1}$. This numerical solution is identical to the exact solution of Eq. (28), because the diagonal of the mesh (Fig. 1) coincides with the characteristic line of the advection equation when $v = a\Delta t/\Delta x = 1$.

Therefore, when the CFL number is either zero or unity, the upwind CESE scheme is not only stable but also exact. Based on the stability analysis above, the remainder of this section will consider only $0 < v < 1$.

### B. Modified equation

To derive the underlying modified equation of the discrete equation (Eq. (31)), a vector function that is sufficiently smooth

$$\boldsymbol{q}(x,t) \equiv \begin{bmatrix} u(x,t) \\ g(x,t) \end{bmatrix} \tag{35}$$

is assumed to satisfy the following evolution equation:

$$\boldsymbol{q}(x, t+\Delta t/2) = \boldsymbol{Q}_L \boldsymbol{q}(x-\Delta x/2, t) + \boldsymbol{Q}_R \boldsymbol{q}(x+\Delta x/2, t) , \tag{36}$$

which is a generalization of Eq. (31). The next step involves converting Eq. (36) into a partial differential form. In this subsection, a constant value of $v$ ($0 < v < 1$) is assumed as $\Delta x$ and $\Delta t$ approach zero, which means $\Delta x$ and $\Delta t$ are infinitesimals of the same order. A Taylor expansion with respect to $x$ and $t$ results in

$$
\begin{aligned}
&\boldsymbol{q}(x,t)+\left(\frac{\Delta t}{2}\right)\boldsymbol{q}_t(x,t)+\frac{1}{2}\left(\frac{\Delta t}{2}\right)^2\boldsymbol{q}_{tt}(x,t)+\frac{1}{6}\left(\frac{\Delta t}{2}\right)^3\boldsymbol{q}_{ttt}(x,t)+O(\Delta t^4)\\
&=\boldsymbol{Q}_L\left[\boldsymbol{q}(x,t)-\left(\frac{\Delta x}{2}\right)\boldsymbol{q}_x(x,t)+\frac{1}{2}\left(\frac{\Delta x}{2}\right)^2\boldsymbol{q}_{xx}(x,t)-\frac{1}{6}\left(\frac{\Delta x}{2}\right)^3\boldsymbol{q}_{xxx}(x,t)+O(\Delta x^4)\right]\\
&+\boldsymbol{Q}_R\left[\boldsymbol{q}(x,t)+\left(\frac{\Delta x}{2}\right)\boldsymbol{q}_x(x,t)+\frac{1}{2}\left(\frac{\Delta x}{2}\right)^2\boldsymbol{q}_{xx}(x,t)+\frac{1}{6}\left(\frac{\Delta x}{2}\right)^3\boldsymbol{q}_{xxx}(x,t)+O(\Delta x^4)\right].
\end{aligned}
\tag{37}
$$

Then, by defining the two matrices $\boldsymbol{A}$ and $\boldsymbol{B}$

$$
\boldsymbol{A}\equiv\boldsymbol{Q}_L+\boldsymbol{Q}_R=\begin{bmatrix}1 & 0\\ 0 & 2\nu-1\end{bmatrix},\qquad \boldsymbol{B}\equiv\boldsymbol{Q}_L-\boldsymbol{Q}_R=\begin{bmatrix}\nu & 1-\nu^2\\ \nu-1 & \nu(2-\nu)\end{bmatrix},
\tag{38}
$$

Eq. (37) can be written as

$$
48\left(\boldsymbol{I}-\boldsymbol{A}\right)\boldsymbol{q}+24\left(\Delta t\boldsymbol{q}_t+\Delta x\boldsymbol{B}\boldsymbol{q}_x\right)=6\left(\Delta x^2\boldsymbol{A}\boldsymbol{q}_{xx}-\Delta t^2\boldsymbol{q}_{tt}\right)-\left(\Delta x^3\boldsymbol{B}\boldsymbol{q}_{xxx}+\Delta t^3\boldsymbol{q}_{ttt}\right)+O(\Delta t^4,\Delta x^4)\,.
\tag{39}
$$

In accordance with the convention of a modified-equation analysis, the temporal derivatives $\boldsymbol{q}_{tt}$ and $\boldsymbol{q}_{ttt}$ in Eq. (39) can be eliminated by utilizing Eq. (39) itself repeatedly. Such an elimination process is essentially the same as the one described by Warming and Hyett;[43] however, the current approach requires more complicated algebraic manipulations. The result is

$$
\boldsymbol{K}\boldsymbol{q}+\boldsymbol{L}\boldsymbol{q}_t+\boldsymbol{M}\boldsymbol{q}_x=\boldsymbol{N}\boldsymbol{q}_{xx}+\boldsymbol{R}\boldsymbol{q}_{xxx}+O(\Delta t^4,\Delta x^4)\,,
\tag{40}
$$

where the coefficient matrices are calculated as

$$
\boldsymbol{K}=48\left(\boldsymbol{I}-\boldsymbol{A}\right)=96\begin{bmatrix}0 & 0\\ 0 & 1-\nu\end{bmatrix},
\tag{41}
$$

$$
\boldsymbol{L}=8\Delta t\left[3\boldsymbol{I}-\left(2\boldsymbol{I}+\boldsymbol{A}\right)\left(\boldsymbol{I}+\boldsymbol{A}\right)^{-1}\left(\boldsymbol{I}-\boldsymbol{A}\right)\right]=8\Delta t\begin{bmatrix}3 & 0\\ 0 & (2\nu^2+2\nu-1)/\nu\end{bmatrix},
\tag{42}
$$

$$
\begin{aligned}
\boldsymbol{M}&=8\Delta x\left\{3\boldsymbol{B}+\left(2\boldsymbol{I}+\boldsymbol{A}\right)\left(\boldsymbol{I}+\boldsymbol{A}\right)^{-1}\boldsymbol{B}\left(\boldsymbol{I}-\boldsymbol{A}\right)\left[\boldsymbol{I}+2\left(\boldsymbol{I}+\boldsymbol{A}\right)^{-1}\left(\boldsymbol{I}-\boldsymbol{A}\right)\right]-\boldsymbol{B}\left(\boldsymbol{I}-\boldsymbol{A}\right)\left(\boldsymbol{I}+\boldsymbol{A}\right)^{-1}\left(\boldsymbol{I}-\boldsymbol{A}\right)\right\}\\
&=8\Delta x\begin{bmatrix}3\nu & -(\nu-1)(\nu+1)(\nu^2-2\nu+4)/\nu\\ 3(\nu-1) & -(\nu-2)(2\nu^2-\nu+2)/\nu\end{bmatrix},
\end{aligned}
\tag{43}
$$

$$
\begin{aligned}
\boldsymbol{N}&=2\Delta x^2\left\{3\boldsymbol{A}-\left(2\boldsymbol{I}+\boldsymbol{A}\right)\left(\boldsymbol{I}+\boldsymbol{A}\right)^{-1}\left[2\boldsymbol{B}^2+\left(2\boldsymbol{B}^2-\boldsymbol{A}\right)\left(\boldsymbol{I}-\boldsymbol{A}\right)+2\boldsymbol{B}\left(\boldsymbol{I}-\boldsymbol{A}\right)\left(\boldsymbol{I}+\boldsymbol{A}\right)^{-1}\boldsymbol{B}\left(3\boldsymbol{I}-\boldsymbol{A}\right)\right]\right.\\
&\left.+\boldsymbol{B}\left[\left(\boldsymbol{I}-\boldsymbol{A}\right)\left(\boldsymbol{I}+\boldsymbol{A}\right)^{-1}\boldsymbol{B}\left(3\boldsymbol{I}-\boldsymbol{A}\right)+\boldsymbol{B}\left(\boldsymbol{I}-\boldsymbol{A}\right)\right]\right\}\\
&=2(\nu-1)\Delta x^2\begin{bmatrix}-(\nu+1)(\nu^2-2\nu+4)/\nu & (\nu+1)(\nu^2-11\nu+16)\\ -(\nu^2+\nu+4)/\nu & -(\nu-2)(5\nu^2+8\nu-1)/\nu\end{bmatrix},
\end{aligned}
\tag{44}
$$

$$
\begin{aligned}
\boldsymbol{R} &= \Delta x^3 \Big\{ -\boldsymbol{B} - (2\boldsymbol{I}+\boldsymbol{A})(\boldsymbol{I}+\boldsymbol{A})^{-1}\Big[2\boldsymbol{B}^3 - \boldsymbol{A}\boldsymbol{B} - \boldsymbol{B}\boldsymbol{A} - 2\boldsymbol{B}(\boldsymbol{I}-\boldsymbol{A})(\boldsymbol{I}+\boldsymbol{A})^{-1}(\boldsymbol{A}-\boldsymbol{B}^2)\Big] \\
&\quad -\boldsymbol{B}\Big[(\boldsymbol{I}-\boldsymbol{A})(\boldsymbol{I}+\boldsymbol{A})^{-1}(\boldsymbol{A}-\boldsymbol{B}^2) - \boldsymbol{B}^2\Big]\Big\} \\
&= (\nu-1)\Delta x^3 \begin{bmatrix} 2(2\nu-3)(\nu+1) & -(\nu+1)(4\nu^2-11\nu+3) \\ (3\nu^3-2\nu^2-6\nu+1)/\nu & -(3\nu^3-5\nu^2-9\nu+5) \end{bmatrix}.
\end{aligned}
\tag{45}
$$

Furthermore, by using the definition of the vector $\boldsymbol{q}$ in Eq. (35), the matrix form of the modified equation (Eq. (40)) can be split into two scalar partial differential equations:

$$
\begin{aligned}
\Delta t u_t + \nu \Delta x u_x &= -\frac{(\nu-1)(\nu+1)(\nu^2-2\nu+4)}{12\nu}\Delta x^2 u_{xx} + \frac{(\nu-1)(\nu+1)(2\nu-3)}{12}\Delta x^3 u_{xxx} \\
&\quad + \frac{(\nu-1)(\nu+1)(\nu^2-2\nu+4)}{3\nu}\Delta x g_x + \frac{(\nu-1)(\nu+1)(\nu^2-11\nu+16)}{12}\Delta x^2 g_{xx} \\
&\quad - \frac{(\nu-1)(\nu+1)(4\nu^2-11\nu+3)}{24}\Delta x^3 g_{xxx} + O(\Delta t^4, \Delta x^4),
\end{aligned}
\tag{46}
$$

$$
\begin{aligned}
g &= \frac{1}{4}\Delta x u_x + \frac{(\nu^2+\nu+4)}{48\nu}\Delta x^2 u_{xx} - \frac{3\nu^3-2\nu^2-6\nu+1}{96\nu}\Delta x^3 u_{xxx} \\
&\quad - \frac{2\nu^2+2\nu-1}{12\nu(1-\nu)}\Delta t g_t + \frac{(\nu-2)(2\nu^2-\nu+2)}{12\nu(1-\nu)}\Delta x g_x + \frac{(\nu-2)(5\nu^2+8\nu-1)}{48\nu}\Delta x^2 g_{xx} \\
&\quad + \frac{3\nu^3-5\nu^2-9\nu+5}{96}\Delta x^3 g_{xxx} + O(\Delta t^4, \Delta x^4).
\end{aligned}
\tag{47}
$$

Recall that the upwind CESE scheme (Eq. (31)) is a numerical approach used to solve the advection equation $u_t + au_x = 0$. Therefore, it is favorable to eliminate the auxiliary function $g(x, t)$ from Eq. (46) and obtain a modified equation for $u(x, t)$. Thus, Eq. (47) is repeatedly used to express the derivatives of $g(x, t)$ in terms of the derivatives of $u(x, t)$. Finally, the desirable form of the modified equation, which does not contain temporal derivatives on its right-hand side, is obtained as follows:

$$
u_t + au_x = \frac{a\Delta x^2}{48}(1+\nu)(1-\nu)u_{xxx} + O(\Delta t^3, \Delta x^3)\,. \tag{48}
$$

The steps to derive Eq. (48) from Eqs. (46) and (47) are outlined in the Appendix.

Based on a comparison between Eq. (48) and the original advection equation $u_t + au_x = 0$, some significant features of the upwind CESE scheme (Eq. (31)) can be inferred. For any given CFL number $\nu$ that satisfies $0 < \nu < 1$, this numerical scheme is second-order accurate, with the leading term of the truncation error represented by

$$
R_3 = \frac{a\Delta x^2}{48}(1+\nu)(1-\nu)u_{xxx} \sim O(\Delta t^2, \Delta x^2)\,. \tag{49}
$$

Ever since the upwind CESE scheme in Eqs. (30)–(32) was proposed, its nominal order of accuracy has been examined only through numerical tests. Here, the modified-equation analysis above provides theoretical evidence for this second-order accuracy. Moreover, Eq. (49) indicates that for a sufficiently small cell size $\Delta x$, the numerical error of this upwind CESE scheme is dominated by positive dispersion, i.e., the leading phase error. When compared to classical schemes for the advection equation, Eq. (49) closely resembles the dominant error term in the modified equation of the Lax–Wendroff scheme, which is given by $(-a\Delta x^2/6)(1 + \nu)(1 - \nu)u_{xxx}$. Notably, the absolute value of the error term (Eq. (49)) for the upwind CESE scheme is only 1/8 of that for the Lax–Wendroff scheme.

### C. Spectral properties

This subsection extends the discussion on the numerical error associated with the present CESE scheme, in terms of its dissipation and dispersion characteristics in wavenumber space.[44–48] For this purpose, it is convenient to consider the initial-value problem of Eq. (28) with the initial condition

$$u(x,0) = e^{\mathrm{i}kx}, \tag{50}$$

which is a representative Fourier mode with wavenumber $k$. This problem can be solved numerically using the upwind CESE scheme represented by Eq. (31), where space and time are discretized as $x_j = j\Delta x$ and $t_n = n\Delta t$. The discrete initial condition for the CESE scheme can be obtained using Eq. (50) and its derivative $u_x(x, 0) = \mathrm{i}ke^{\mathrm{i}kx}$, such that

$$\boldsymbol{q}_j^0 = \begin{bmatrix} u(x_j,0) \\ \dfrac{\Delta x}{4} u_x(x_j,0) \end{bmatrix} = \boldsymbol{A}_0 e^{\mathrm{i}j\varphi}, \quad \boldsymbol{A}_0 \equiv \begin{bmatrix} 1 \\ \mathrm{i}\dfrac{\varphi}{4} \end{bmatrix}, \tag{51}$$

where $\varphi = k\Delta x$ is the reduced wavenumber,[44] and $0 \le \varphi \le \pi$. The solution of the discrete equation (Eq. (31)), with the initial condition from Eq. (51), takes the form $\boldsymbol{q}_j^n = \boldsymbol{A}_n e^{\mathrm{i}j\varphi}$, and therefore, Eq. (31) leads to

$$\boldsymbol{A}_n = \boldsymbol{M}\boldsymbol{A}_{n-1/2}, \quad \boldsymbol{M} \equiv e^{-\mathrm{i}\varphi/2}\boldsymbol{Q}_L + e^{\mathrm{i}\varphi/2}\boldsymbol{Q}_R = \begin{bmatrix} \cos\dfrac{\varphi}{2} - \mathrm{i}\nu\sin\dfrac{\varphi}{2} & \mathrm{i}(\nu^2-1)\sin\dfrac{\varphi}{2} \\ \mathrm{i}(1-\nu)\sin\dfrac{\varphi}{2} & (2\nu-1)\cos\dfrac{\varphi}{2} + \mathrm{i}(\nu^2-2\nu)\sin\dfrac{\varphi}{2} \end{bmatrix}. \tag{52}$$

Subsequently, using Eq. (52) $2n$ times yields $\boldsymbol{A}_n = \boldsymbol{M}^{2n}\boldsymbol{A}_0$, and the solution of the CESE scheme is

$$\boldsymbol{q}_j^n = \boldsymbol{M}^{2n}\boldsymbol{A}_0 e^{\mathrm{i}j\varphi}, \tag{53}$$

where $\boldsymbol{A}_0$ and $\boldsymbol{M}$ are defined in Eqs. (51) and (52), respectively.

The eigenvalues of $\boldsymbol{M}$ can be calculated. They depend on the CFL number $\nu$ and reduced wavenumber $\varphi$:

$$\lambda_{1,2} = \left[\nu\cos\frac{\varphi}{2} + \frac{\mathrm{i}}{2}(\nu^2 - 3\nu)\sin\frac{\varphi}{2}\right] \pm \frac{\sqrt{2}}{4}(1-\nu)\sqrt{(8+4\nu-\nu^2)+(\nu^2-4\nu)\cos\varphi + 4\mathrm{i}\,\nu\sin\varphi}\,. \tag{54}$$

It can be shown that for $0 < \nu < 1$ and $0 \le \varphi < \pi$, the two eigenvalues are not equal and that $|\lambda_1| > |\lambda_2|$. By using these eigenvalues and their corresponding eigenvectors, Eq. (53) can be simplified. Its first row is the solution of $u_j^n$:

$$u_j^n = \left(\frac{\lambda_2 - \lambda_0}{\lambda_2 - \lambda_1}\lambda_1^{2n} + \frac{\lambda_1 - \lambda_0}{\lambda_1 - \lambda_2}\lambda_2^{2n}\right)e^{\mathrm{i}j\varphi}, \tag{55}$$

where the parameter $\lambda_0 = \cos(\varphi/2) + (1-\nu^2)(\varphi/4)\sin(\varphi/2) - \mathrm{i}\,\nu\sin(\varphi/2)$. Equation (55) can be regarded as an analytical solution of the discrete equation of the CESE scheme. The difference between Eq. (55) and the exact solution of the original differential equation is a measure of the error associated with the CESE discretization.

The exact solution $u(x,t) = e^{\mathrm{i}k(x-at)}$ for Eq. (28) with the initial condition (Eq. (50)) can be evaluated at the mesh point as

$$u(x_j, t_n) = e^{\mathrm{i}k(j\Delta x - an\Delta t)} = e^{-\mathrm{i}n\nu\varphi}e^{\mathrm{i}j\varphi}\,. \tag{56}$$

For comparison, any discrete solution of this initial-value problem is expected to be written in the form

$$u_j^n = e^{-\mathrm{i}n\nu\Phi}e^{\mathrm{i}j\varphi}, \tag{57}$$

where $\Phi$ is the modified wavenumber.[44] In fact, owing to the coexistence of two eigenvalues, the CESE solution in Eq. (55) differs from Eq. (57) in a general sense. Nevertheless, the modified-wavenumber analysis can proceed in the limiting case of $n\to\infty$. Because $|\lambda_1| > |\lambda_2|$ for given parameters $0 < \nu < 1$ and $0 \le \varphi < \pi$, the effect of $\lambda_2$ on Eq. (55) diminishes with the growth of $n$, and for a sufficiently large $n$, the CESE solution becomes dominated solely by the first term in Eq. (55). Hence, a comparison of Eqs. (55) and (57) shows that $\Phi \to (2\mathrm{i}/\nu)\ln\lambda_1$, when $n$ approaches infinity. By using the expression of $\lambda_1$ in Eq. (54), the modified wavenumber for this upwind CESE scheme can be calculated as

$$\Phi(\varphi,\nu) = \frac{2\mathrm{i}}{\nu}\ln\left\{\left[\nu\cos\frac{\varphi}{2} + \frac{\mathrm{i}}{2}(\nu^2 - 3\nu)\sin\frac{\varphi}{2}\right] + \frac{\sqrt{2}}{4}(1-\nu)\sqrt{(8+4\nu-\nu^2)+(\nu^2-4\nu)\cos\varphi + 4\mathrm{i}\,\nu\sin\varphi}\right\}. \tag{58}$$

The real and imaginary parts of this modified wavenumber, as functions of the reduced wavenumber $\varphi$ and the CFL number $\nu$, are illustrated in Fig. 5 and Fig. 6, respectively. The exact solution (Eq. (56)) corresponds to $\mathrm{Re}[\Phi] = \varphi$ and $\mathrm{Im}[\Phi] = 0$, which are also plotted as benchmarks in these figures.

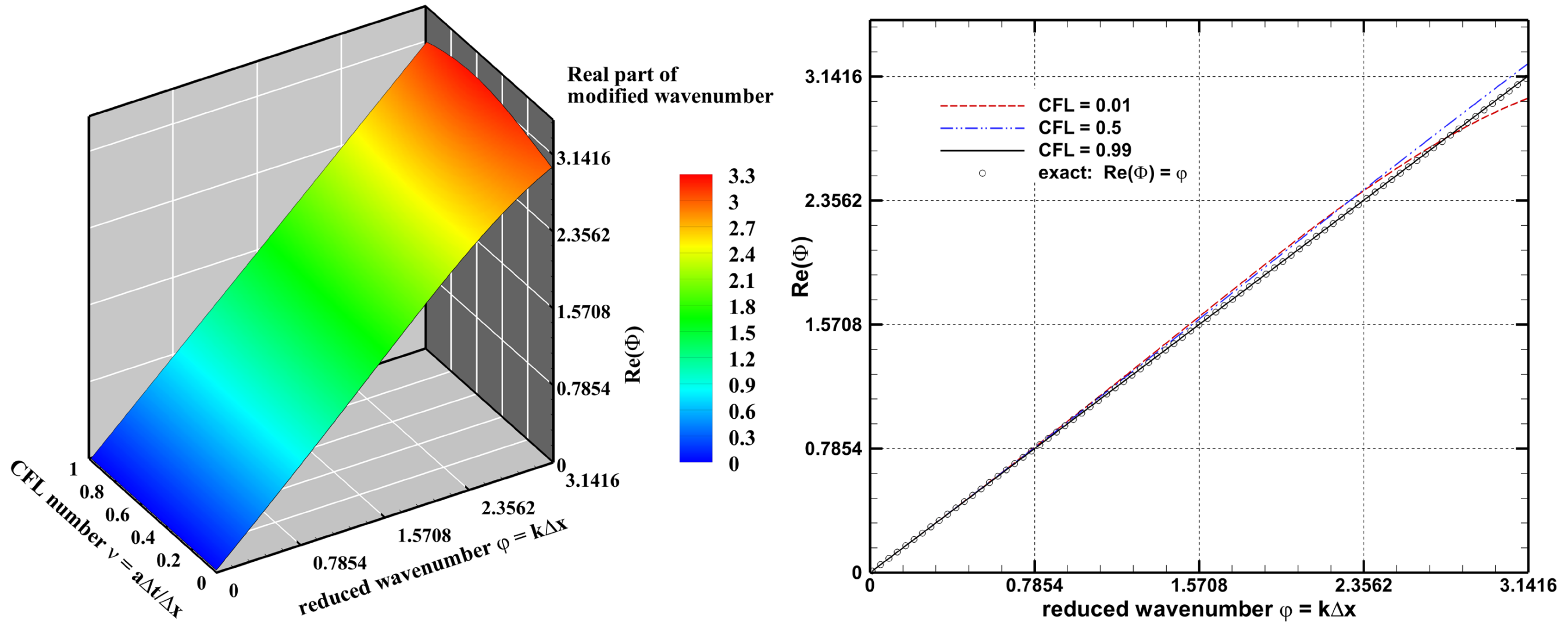

FIG. 5. Real part of modified wavenumber. Left: Surface of Re[$\Phi(\varphi, \nu)$]. Right: Curves of Re[$\Phi(\varphi)$] for different values of $\nu$.

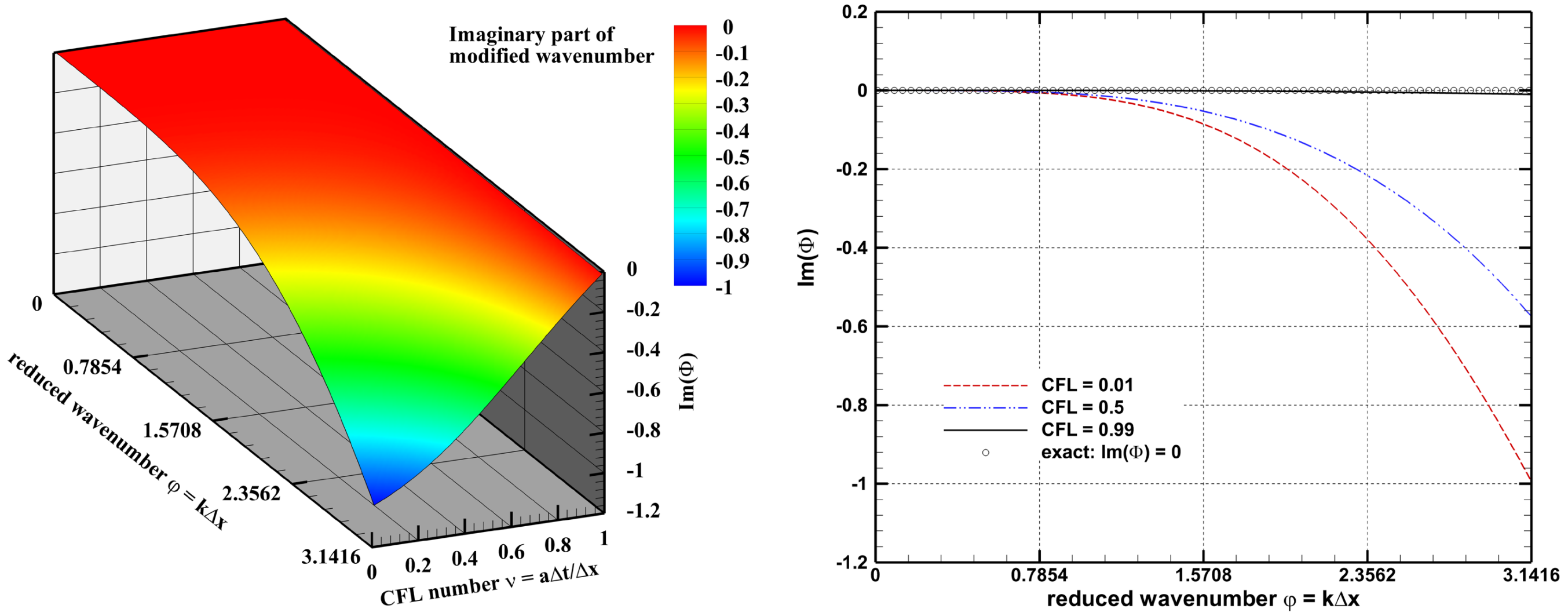

FIG. 6. Imaginary part of modified wavenumber. Left: Surface of Im[$\Phi(\varphi, \nu)$]. Right: Curves of Im[$\Phi(\varphi)$] for different values of $\nu$.

The implication of the modified wavenumber can be elucidated by rewriting Eq. (57) as follows:

$$u_j^n = (e^{\nu\,\mathrm{Im}[\Phi]})^n \exp\left\{ik\left[x_j - \left(\mathrm{Re}[\Phi]/\varphi\right)at_n\right]\right\}. \tag{59}$$

Using this, Re[$\Phi$] /$\varphi$ can be identified as the ratio of the numerical phase speed $a_p(\varphi, \nu)$ to the exact wave speed $a$, and $e^{\nu\mathrm{Im}\Phi}$ as the decay factor of the wave amplitude for one time step. The numerical dispersion characteristics of the upwind CESE scheme can be demonstrated by $a_p(\varphi, \nu)/a$ = Re[$\Phi$] /$\varphi$, as shown in Fig. 7. The dissipation characteristics are plotted in Fig. 8, which depicts $A(\varphi, \nu) = e^{\nu\mathrm{Im}[\Phi]}$.

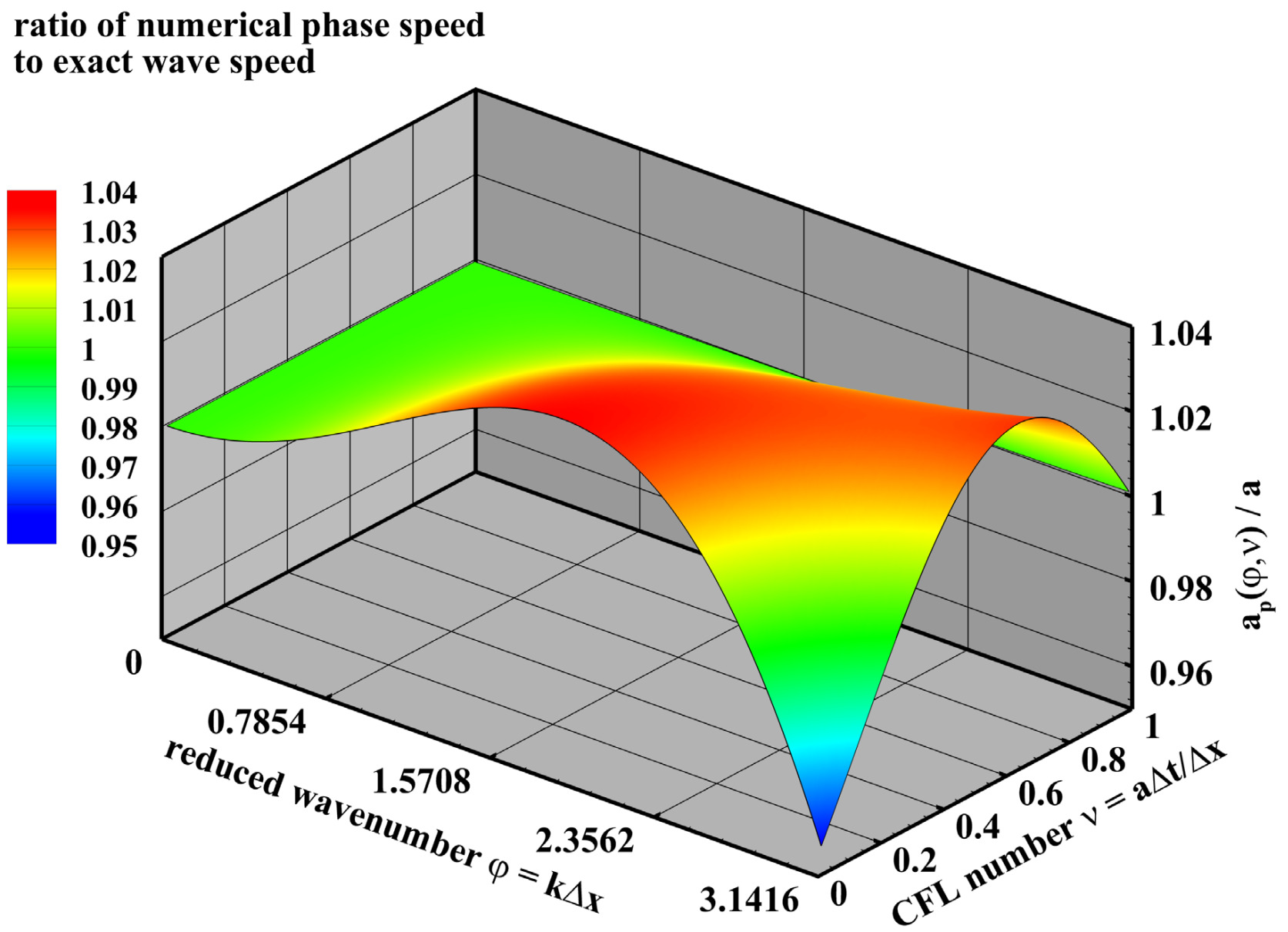

FIG. 7. Numerical dispersion of the upwind CESE scheme as a function of the reduced wavenumber $\varphi = k\Delta x$ and the CFL number $\nu = a\Delta t/\Delta x$. The surface represents the ratio of the numerical phase speed $a_{\mathrm{p}}(\varphi, \nu)$ to the exact wave speed $a$, and $a_{\mathrm{p}}(\varphi, \nu)/a = \mathrm{Re}[\Phi]/\varphi$.

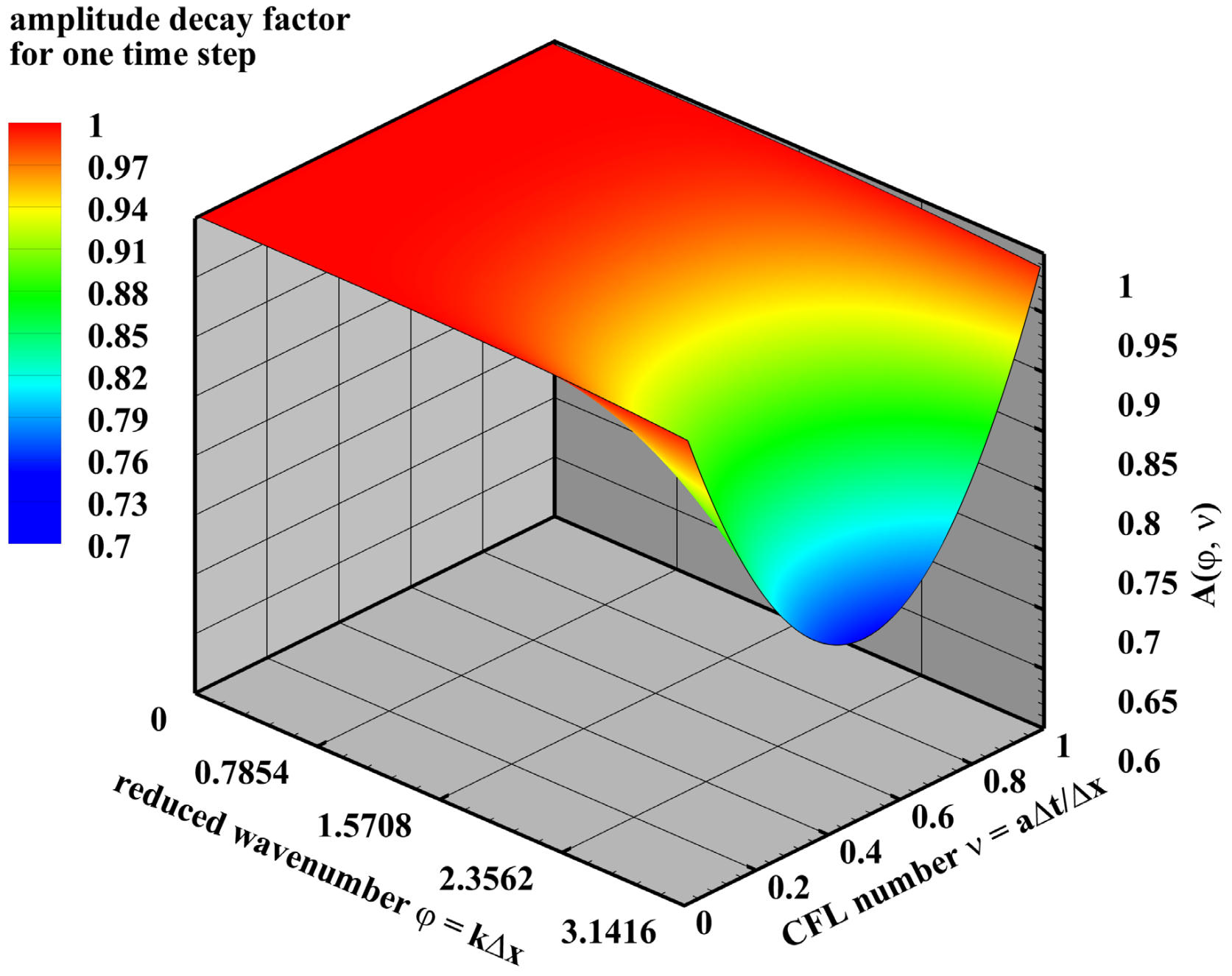

FIG. 8. Numerical dissipation of the upwind CESE scheme as a function of the reduced wavenumber $\varphi = k\Delta x$ and the CFL number $\nu = a\Delta t/\Delta x$. The surface represents the decay factor of the wave amplitude for one time step, which is $A(\varphi, \nu) = e^{\nu \mathrm{Im}[\Phi]}$.

## IV. NUMERICAL EXAMPLES

In this section, eight different numerical examples are considered to demonstrate the performance of the upwind CESE scheme. For this purpose, in previous studies,[10,20,27,28] some basic benchmark problems have been used to test the upwind CESE scheme. The convergence rate tests, the low-density Riemann problem,[49] and the slowly moving or stationary contact discontinuity problem are detailed in Ref. 27. Various 2D Riemann problems[50] and the double Mach reflection problem[51] are detailed in Ref. 28. Sedov's blast wave problem[52] and the viscous shock tube problem[53] are presented in Ref. 20. Sod's shock-tube problem[54] and the Mach 3 forward-facing step problem[51] are presented in Ref. 10. The examples in this section are intended to complement the examination of the upwind CESE scheme without repeating the above-mentioned test problems. The performance of the scheme in this section serves to support the findings in Sec. III. The theoretical analysis in Sec. III, in turn, explains the capabilities of the upwind CESE scheme for these examples.

### A. Sine waves

First, the upwind CESE scheme is applied to the linear advection of sine waves with different wavenumbers. The governing equation (Eq. (28)) is solved, in which $-1 \le x \le 1$, $0 \le t < +\infty$, and the wave speed $a = 1$. The periodic boundary condition is implemented at $x = -1$ and $x = 1$. The initial condition is

$$u(x,0) = \sin(kx), \qquad -1 \le x \le 1, \tag{60}$$

where the wavenumber $k$ is set to three different values: $\pi$, $10\pi$, and $25\pi$.

In the CESE computation, the computational domain is divided into 200 cells, with a uniform cell size of $\Delta x = 0.01$. This results in different values of $k$ corresponding to different reduced wavenumbers $\varphi = k\Delta x$, which are equal to $\pi/100$, $\pi/10$, and $\pi/4$, respectively. The time step size is determined by the CFL number $\nu = a\Delta t/\Delta x$, and three different values ($\nu = 0.01$, 0.50, and 1.00) are considered. Therefore, a total of nine runs are conducted. In each run, the numerical results of $u(x, t)$ are output at time $t = 2$ and then compared with the exact solution of the advection equation.

In each row of Fig. 9, the results of simulations at a specific CFL number for different reduced wavenumbers are presented. For all values of $\nu$ (0.01, 0.50, and 1.00), the simulations accurately capture sine waves fluctuating at low and medium wavenumbers. However, high-wavenumber fluctuations are considerably dampened in the simulations using $\nu = 0.01$ or 0.50. This observation is consistent with the analysis of the dissipation rate of the upwind CESE scheme, as shown in Fig. 6. As the CFL number approaches unity, the numerical dissipation and dispersion diminish, and the exact solution is recovered.

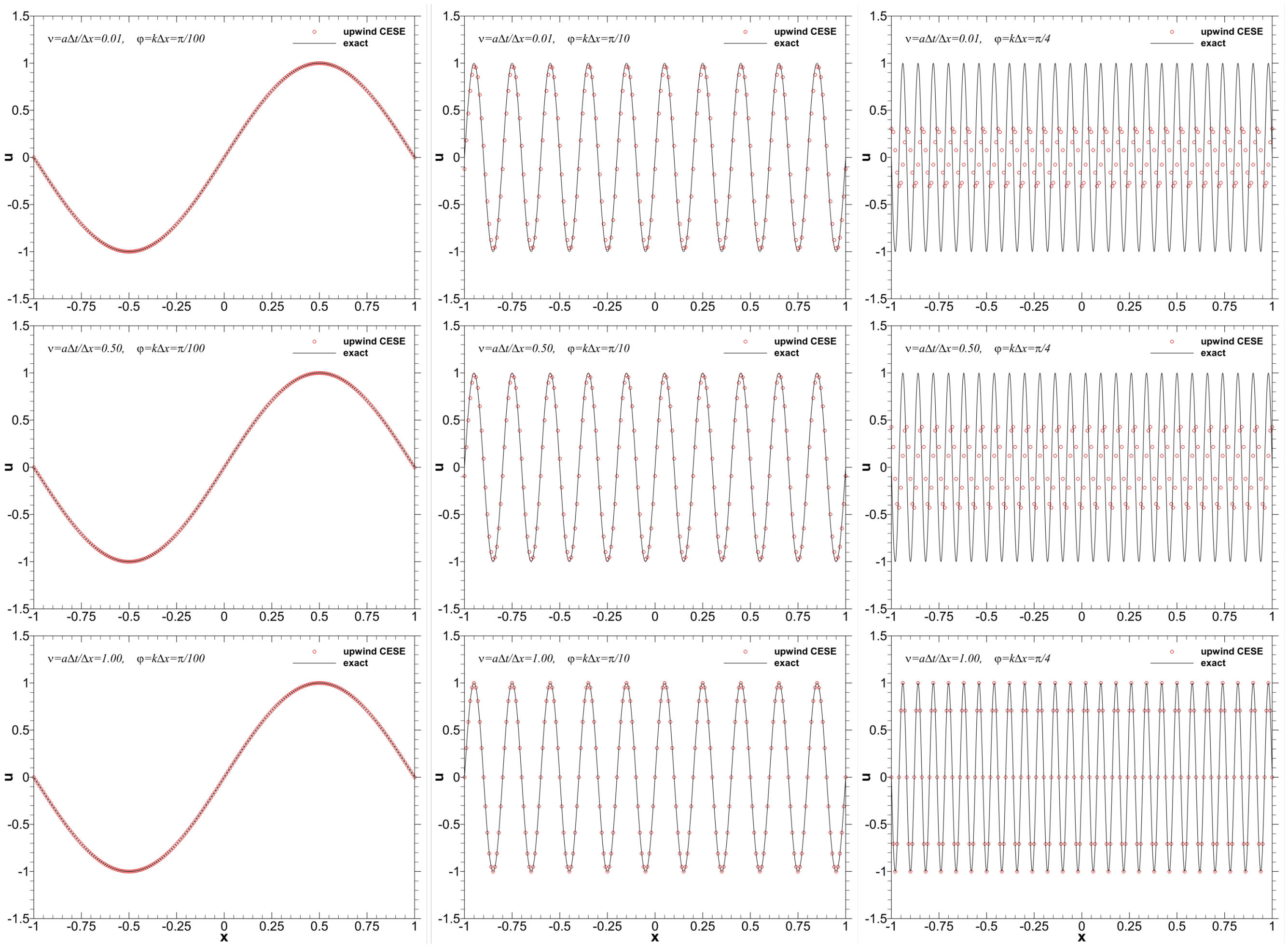

FIG. 9. Advection of sine waves. Left, middle, and right columns: the reduced wavenumber $\varphi = k\Delta x = \pi/100$, $\pi/10$, and $\pi/4$, respectively. Top, middle, and bottom rows: the CFL number $\nu = a\Delta t/\Delta x = 0.01$, 0.50, and 1.00, respectively.

## B. Multiple waves

This example, designed by Jiang and Shu,[55] is widely used to test the numerical scheme for solving the initial-value problem of the linear advection equation (Eq. (28)). Here, the wave speed is $a = 1$, and the spatial domain $[-1, 1]$ is considered. The periodic boundary condition is implemented at $x = -1$ and $x = 1$. At time $t = 0$, the initial condition is

$$u(x,0)=\begin{cases}\left[G(x,\beta,z-\delta)+G(x,\beta,z+\delta)+4G(x,\beta,z)\right]/6, & -0.8\le x\le -0.6,\\ 1, & -0.4\le x\le -0.2,\\ 1-\left|10(x-0.1)\right|, & 0\le x\le 0.2,\\ \left[F(x,\alpha,a-\delta)+F(x,\alpha,a+\delta)+4F(x,\alpha,a)\right]/6, & 0.4\le x\le 0.6,\\ 0, & \text{otherwise},\end{cases} \tag{61}$$

$$G(x,\beta,z)=e^{-\beta(x-z)^2}, \qquad F(x,\alpha,a)=\sqrt{\max\left\{1-\alpha^2(x-a)^2,\ 0\right\}}.$$

The parameters in Eq. (61) are taken as $z = -0.7$, $a = 0.5$, $\alpha = 10$, $\delta = 0.005$, and $\beta = \ln 2/(36\delta^2)$.

The CESE computation is performed with a uniform mesh consisting of 200 cells. The corresponding cell size is $\Delta x = 0.01$, and the time step size is determined by the CFL number $\nu = a\Delta t/\Delta x = 0.88$. At time $t = 8$, the numerical solution of $u(x, t)$ is output and compared with the exact solution of the advection equation, as shown in Fig. 10. Both solutions depict the advection of a combination of Gaussian, square, triangular, and elliptical waves. Satisfactory accuracy is achieved with the current 2nd-order scheme and computational mesh. The numerical results are free from spurious oscillations or overshoots.

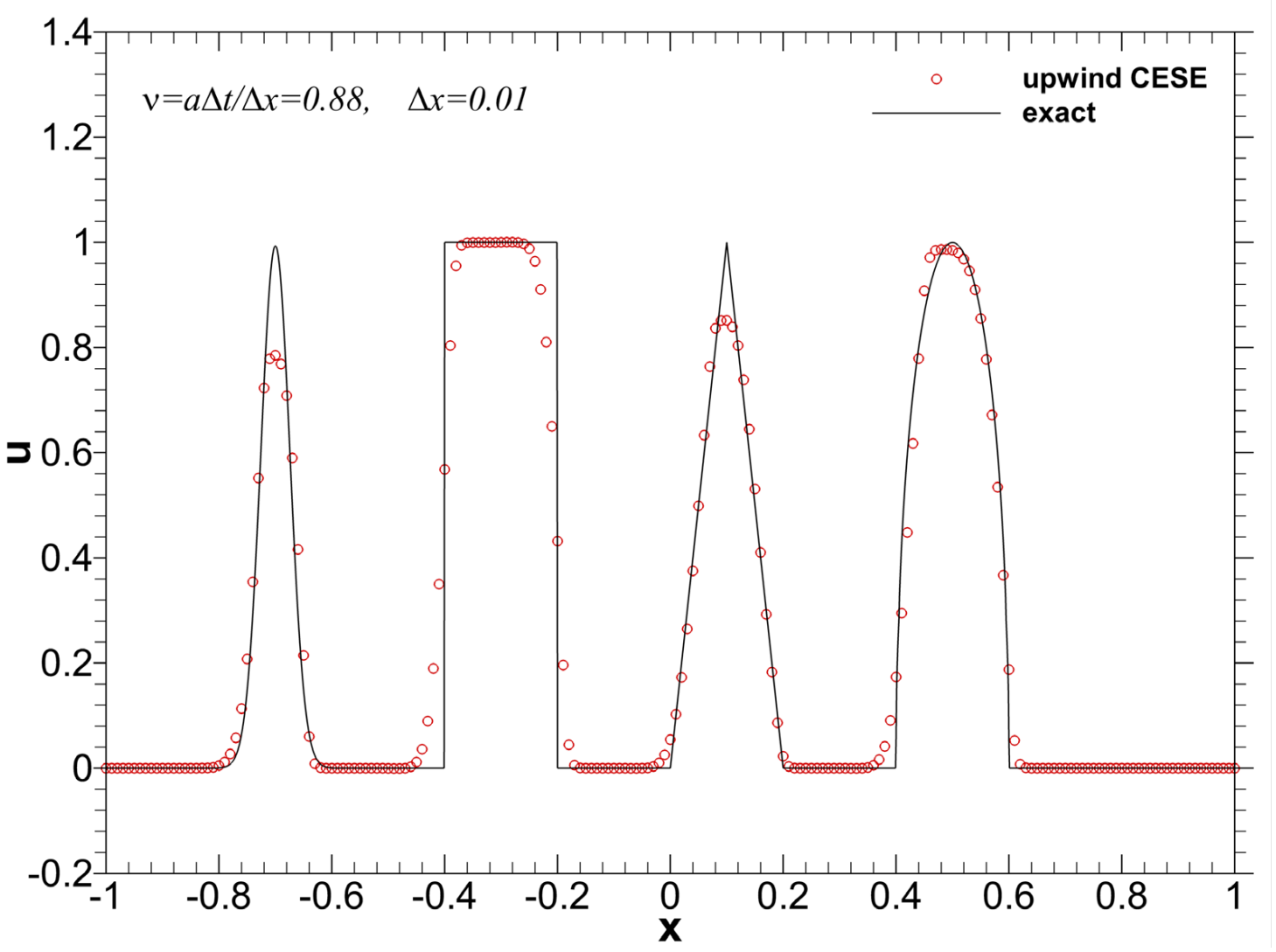

FIG. 10. Linear advection of multiple waves. The number of mesh cells is 200 and the solution is shown at time $t = 8$.

## C. Woodward–Colella problem

The problem of two interacting blast waves, as proposed by Woodward and Colella,[51] is used to validate the robustness and accuracy of numerical schemes. The governing equation considered here is the 1D unsteady compressible Euler equation (Eq. (26)) for a calorically perfect gas ($\gamma = 1.4$), in which $0 \le x \le 1$. The boundary conditions at $x = 0$ and $x = 1$ are assumed to be reflective (solid walls). The initial condition at time $t = 0$ involves two discontinuities and three constant states as follows:

$$(\rho,\ u,\ p) = \begin{cases} (1,\ 0,\ 1000), & 0.0 \le x < 0.1, \\ (1,\ 0,\ 0.01), & 0.1 \le x < 0.9, \\ (1,\ 0,\ 100), & 0.9 \le x \le 1.0. \end{cases} \tag{62}$$

In the CESE computation, the computational domain [0, 1] is discretized by a uniform mesh of 800 cells. The time evolution of the gas system is simulated with a CFL number of 0.8, and the flow field at $t = 0.038$ is displayed in Fig. 11. The reference solution was obtained by Woodward and Colella[51] using a carefully designed mesh with local refinement. Complex interactions between multiple strong waves pose challenges to the stability of the scheme as well as its resolution. In the CESE solution, the discontinuities are captured sharply. Meanwhile, the peaks and valleys of the density distribution are well resolved.

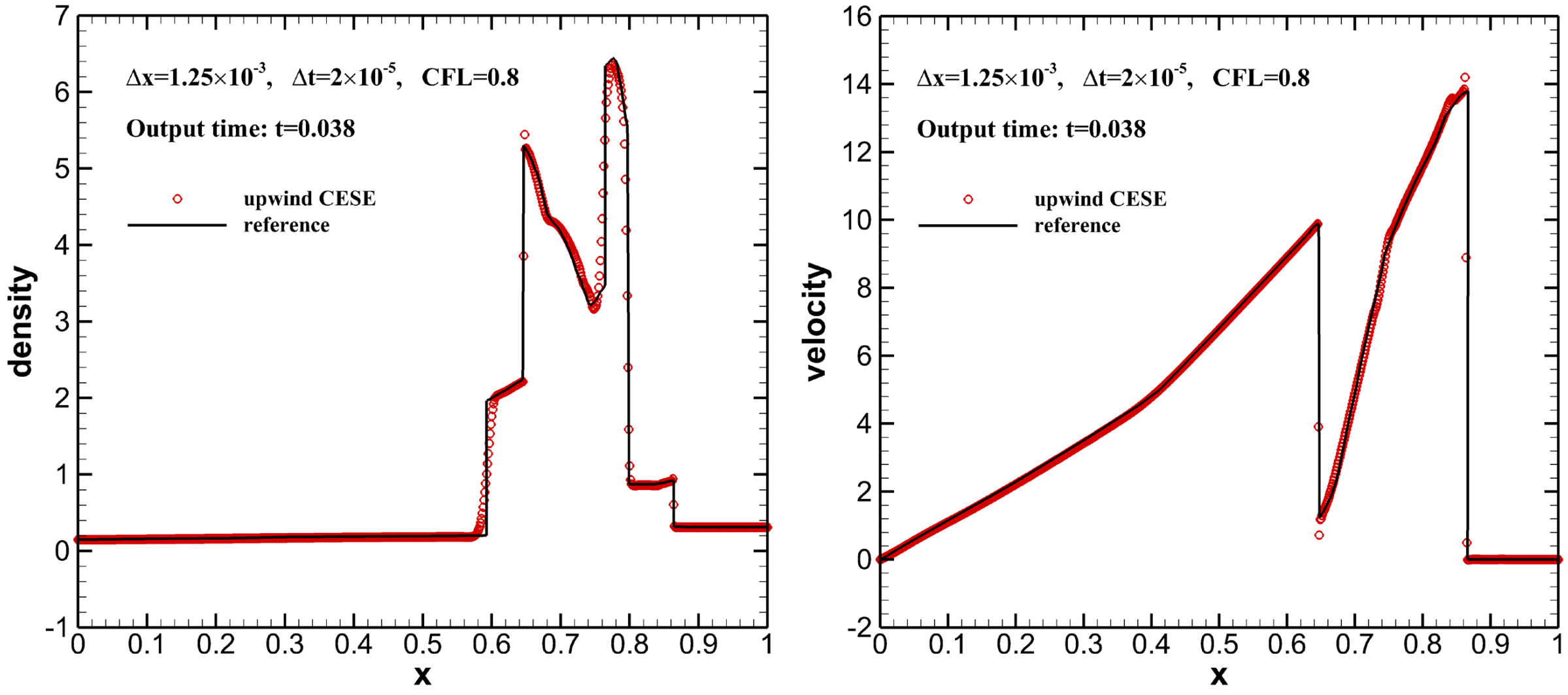

FIG. 11 The distributions of density (left) and velocity (right) after the collision of two strong blast waves. The upwind CESE results are obtained on a uniform mesh of 800 cells. The solution in Ref. 51 serves as the benchmark.

## D. Shu–Osher problem

This case was proposed by Shu and Osher.[56] It is characterized by the presence of both discontinuities and fine structures in smooth regions. The 1D unsteady compressible Euler equation (Eq. (26)) for a calorically perfect gas ($\gamma = 1.4$) is solved on the domain of $0 \le x \le 10$. The non-reflecting boundary condition is imposed at $x = 0$ and $x = 10$. The initial flow field is

$$(\rho,\ u,\ p) = \begin{cases} (3.857143,\ 2.629369,\ 10.33333), & 0 \le x < 1, \\ (1 + 0.2\sin(5x),\ 0,\ 1), & 1 \le x \le 10, \end{cases} \tag{63}$$

which means a perturbation in density is added to the pre-shock field of a Mach 3 traveling shock wave.

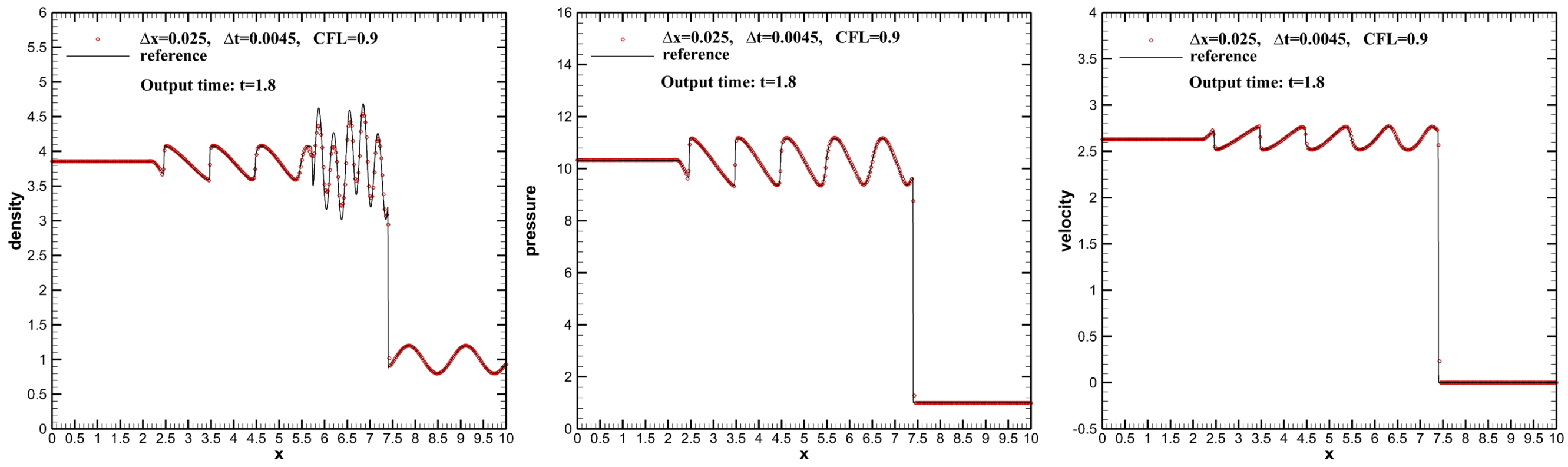

FIG. 12. CESE results of the Shu–Osher problem on a uniform mesh of 400 cells. Left: density. Middle: pressure. Right: velocity.

The upwind CESE scheme is used to solve this problem using 400 mesh cells of equal size and a CFL number of 0.9. As shown in Fig. 12, the distributions of density, pressure, and velocity at time $t = 1.8$ are examined and compared with a reference solution. This reference solution is generated using the 5th-order WENO scheme[55] with 4000 mesh cells of equal size and a CFL number of 0.25.

The shock wave interacts with the fluctuating density field, resulting in a complex flow with multi-scale structures. Considering the limited number of mesh cells and the fact that the current scheme is 2nd-order, the upwind CESE results demonstrate excellent accuracy and reasonable agreement with the reference solution. The shock waves are accurately captured, and the post-shock high-wavenumber fluctuations in density are also well resolved.

## E. Kelvin–Helmholtz instability

The Kelvin–Helmholtz instability is a notable phenomenon that occurs at interfaces in shear flows. In this study, an inviscid shear flow, as described in Ref. 57, is simulated by solving the 2D unsteady compressible Euler equation for a calorically perfect gas ($\gamma = 5/3$). The computational domain spans [0, 1] in both the $x$- and $y$- directions. Periodic boundary conditions are enforced on all four boundaries. The initial condition at $t = 0$ is designed as follows. Within the central horizontal strip defined by $0.25 < y < 0.75$, the density and the horizontal velocity of the fluid are $\rho = 2$ and $u = 0.5$, respectively. In the rest of the domain, $\rho = 1$ and $u = -0.5$. The pressure $p = 2.5$ remains constant across the entire domain. To induce a single-mode instability, the vertical velocity $v$ is perturbed as follows:

$$v(x, y) = w_0 \sin(4\pi x)\left\{\exp\left[-\frac{(y-0.25)^2}{2\sigma^2}\right] + \exp\left[-\frac{(y-0.75)^2}{2\sigma^2}\right]\right\}, \tag{64}$$

where $w_0 = 0.1$ and $\sigma^2 = 1.25\times10^{-3}$. This perturbation is concentrated primarily near the two interfaces of the shear flow.

Two different uniform Cartesian meshes, consisting of 500×500 and 1000×1000 cells, are used in this study. The simulation results are shown in Fig. 13. They highlight the effectiveness of the upwind CESE scheme in accurately capturing fine structures induced by the shearing process. As the resolution is increased, the numerical dissipation is reduced and more small-scale structures resulting from the Kelvin–Helmholtz instability can be evidently resolved.

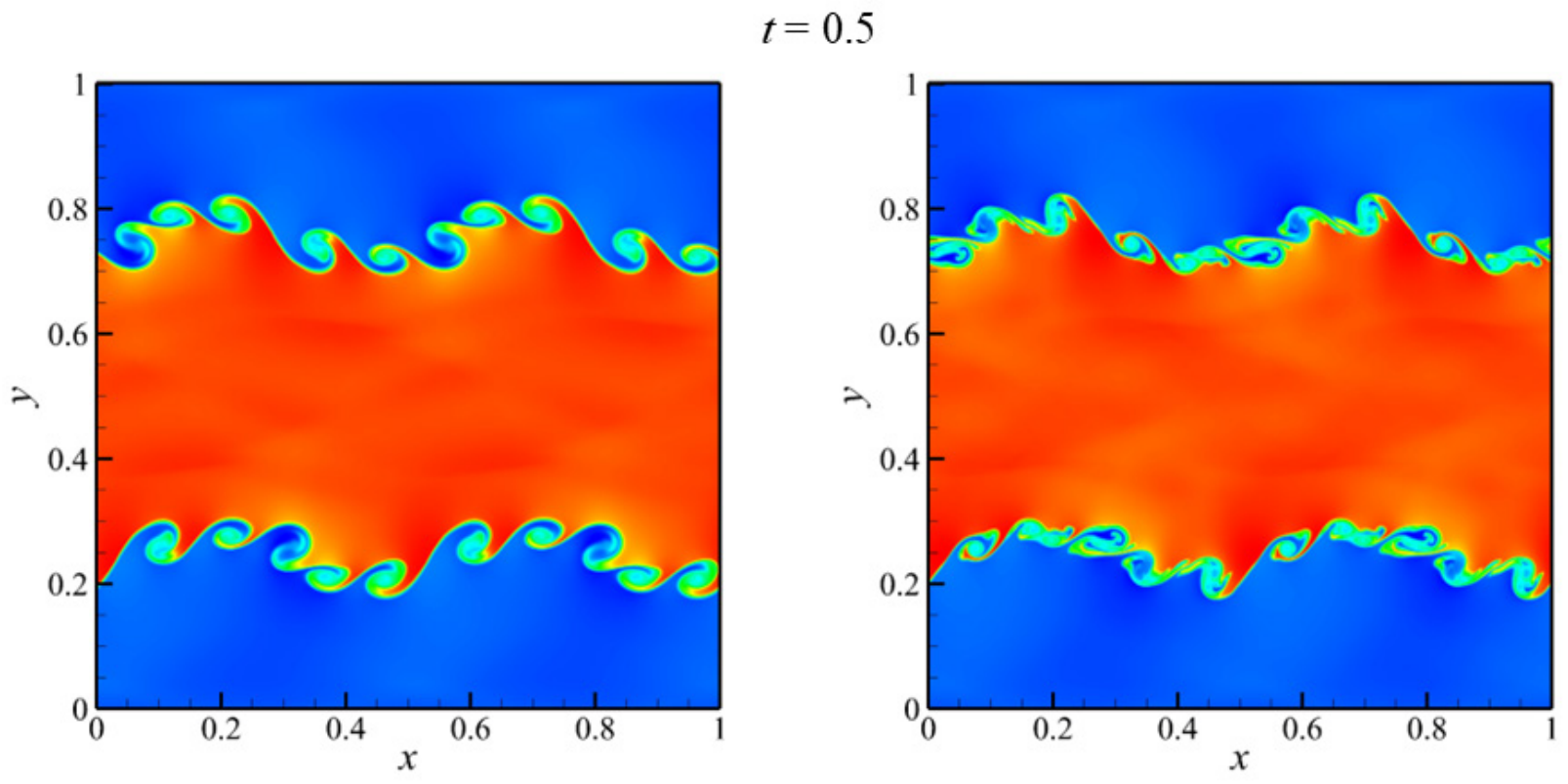

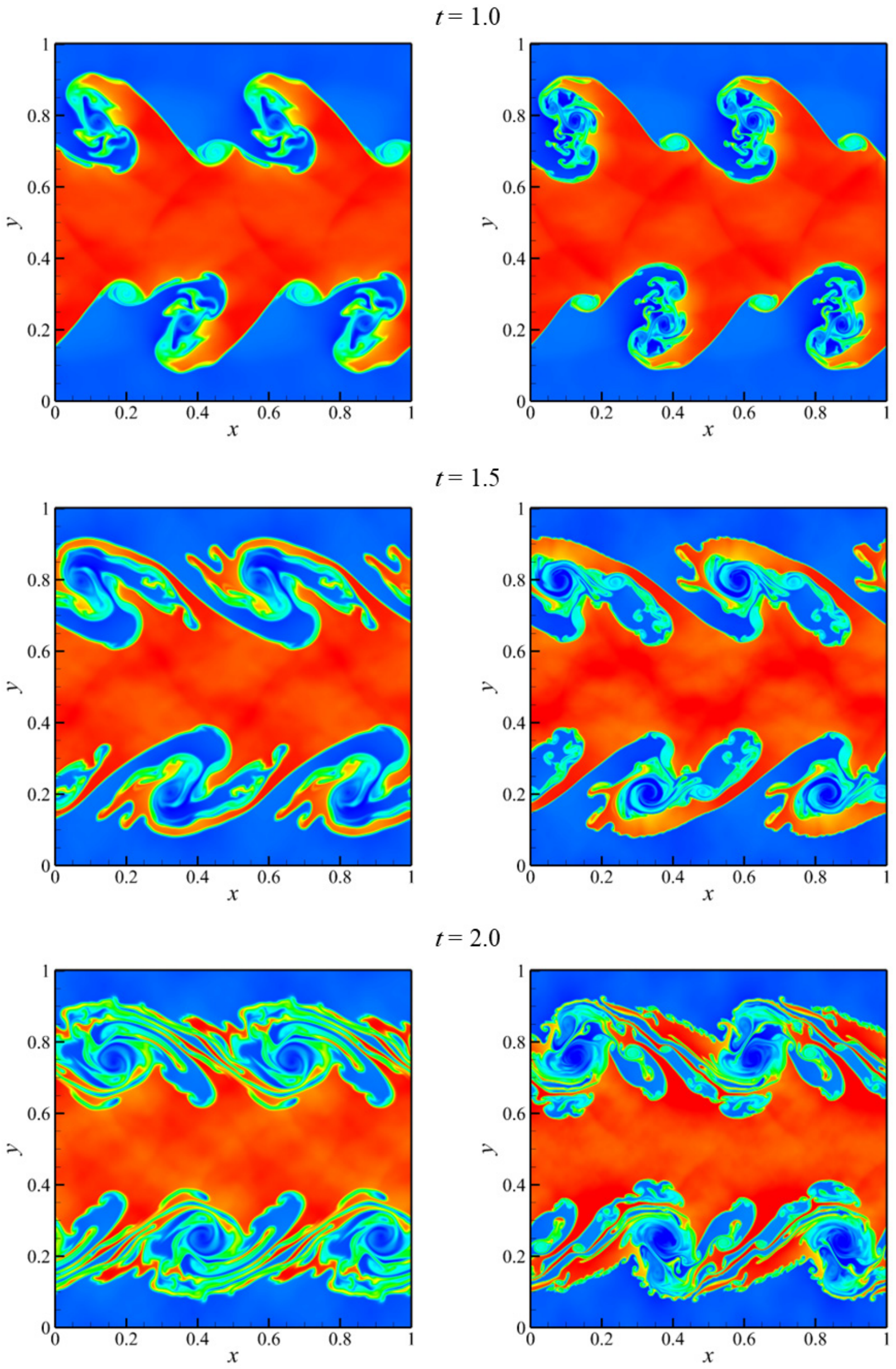

FIG. 13. Density contours of the Kelvin–Helmholtz instability problem at four different times. Left: number of mesh cells is 500×500; right: number of mesh cells is 1000×1000.

## F. 2D isentropic vortex

To verify the order of accuracy of the upwind CESE scheme in simulating two-dimensional flows, a test of the evolution of a 2D isentropic vortex[58] is conducted. The gas dynamics are governed by the compressible Euler equation ($\gamma = 1.4$) on the domain $[-5, 5] \times [-5, 5]$. The background flow is characterized by the following parameters: $\rho = 1$, $p = 1$, $u = 1$, and $v = 1$. Initially, an isentropic vortex is superimposed upon the background flow. It is generated by perturbations in velocity ($u$, $v$) and temperature $T = p/\rho$, while maintaining a constant entropy $S = p/\rho^{\gamma}$. These perturbations can be expressed as follows:

$$(\delta u,\ \delta v) = \frac{\chi}{2\pi} e^{0.5(1-r^2)} (-y,\ x),$$
$$\delta T = -\frac{(\gamma - 1)\chi^2}{8\gamma\pi^2} e^{1-r^2}, \tag{65}$$
$$\delta S = 0,$$

where $r = (x^2 + y^2)^{1/2}$ represents the radial distance from the origin, and $\chi = 5$ denotes the vortex strength. Periodic boundary conditions are enforced at all boundaries.

The exact solution for the above problem is the passive convection of the vortex at the background flow velocity. To assess the order of accuracy of the upwind CESE scheme, five different uniform Cartesian meshes with mesh cell sizes of $\Delta h = 1$, 1/2, 1/4, 1/8, and 1/16 are used in the computations. Comparisons are made between the numerical results and the exact solution at time $t = 2.0$. The $L^1$ and $L^\infty$ errors, calculated in terms of the density field $\rho$, are plotted against the cell size $\Delta h$ in Fig. 14. The results indicate that, in this case, the upwind CESE scheme demonstrates slightly higher than second-order accuracy.

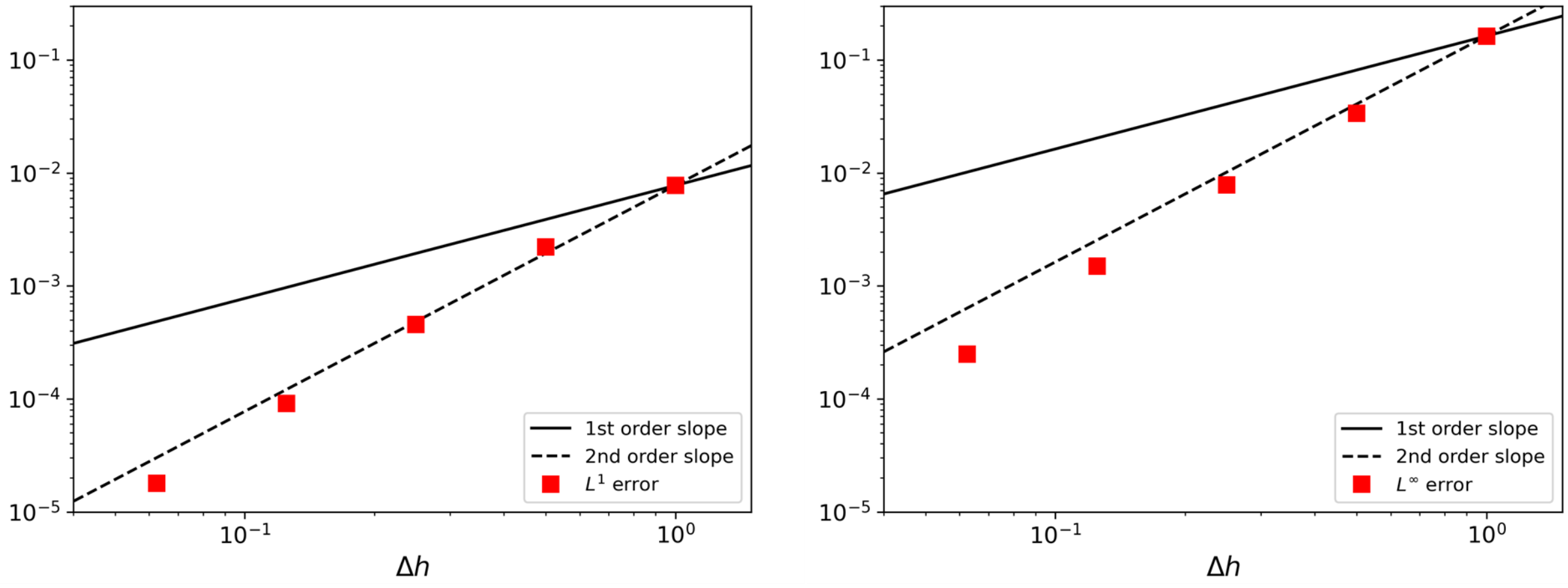

FIG. 14. Results of the accuracy test for the 2D isentropic vortex problem. Left: $L^1$ error; right: $L^\infty$ error. Solid and dashed lines have slopes of 1 and 2 in logarithmic coordinates, respectively.

## G. Shock–vortex interaction

In this subsection, the strong interaction between a vortex and a shock wave[59–63] is studied. This interaction involves a combination of smooth flow structures and multiple discontinuities. This particular case has also been analyzed in the context of aeroacoustic noise.[60] The dynamics are assumed to be governed by the compressible Euler equation for a perfect gas, with a specific heat ratio of $\gamma = 1.4$. The rectangular computational domain spans [0, 2] in the $x$-direction and [0, 1] in the $y$-direction.

Initially, a stationary shock wave with a Mach number $M_s = 1.5$ is located at $x = 0.5$. The flow properties on the left side of the shock are defined as $\rho_1 = 1$, $u_1 = \sqrt{\gamma} M_s$, $v_1 = 0$, and $p_1 = 1$. The initial conditions on the right side of the shock are calculated using the Rankine–Hugoniot relations. Additionally, a vortex is positioned with its center at (0.25, 0.50), which rotates counterclockwise with the tangential velocity ($v_\theta$) given by

$$v_\theta = \begin{cases} v_m \dfrac{r}{a}, & \text{if } r \le a \\ v_m \dfrac{a}{a^2 - b^2}\left(r - \dfrac{b^2}{r}\right), & \text{if } a < r \le b \\ 0, & \text{if } r > b \end{cases} \tag{66}$$

where $v_m$ represents the maximum tangential velocity, and $r$ indicates the distance from the vortex center. The geometrical parameters are $a = 0.075$ and $b = 0.175$. The strength of the vortex, characterized as $M_v = v_m / \sqrt{\gamma}$, is set to be 0.9. The density and pressure fields follow the isentropic relations:

$$\rho = \rho_1 \left(T/T_1\right)^{\frac{1}{\gamma-1}}, \qquad p = p_1 \left(T/T_1\right)^{\frac{\gamma}{\gamma-1}}. \tag{67}$$

Substituting Eq. (67) into the radial momentum equation results in a differential equation for temperature:

$$\frac{\mathrm{d}T}{\mathrm{d}r} = \frac{\gamma - 1}{\gamma} \frac{v_\theta^2}{r}. \tag{68}$$

Equations (66)–(68) can be used to calculate the pressure and density distributions. The velocity field is determined by the superposition of the tangential velocity and the pre-shock velocity as follows:

$$(u,\ v) = (u_1,\ v_1) + v_\theta(-\sin\theta,\ \cos\theta). \tag{69}$$

A supersonic inlet condition is applied to the left boundary, while a subsonic outlet condition is applied to the right boundary. The upper and lower boundaries are assumed to be reflective. The computations are carried out using three different mesh resolutions: 1000×500 cells, 2000×1000 cells, and 4000×2000 cells. The output time for all computations is $t = 0.7$.

In this strong shock–vortex interaction, the circular vortex undergoes compression by the shock wave, resulting in the formation of elliptical structures. The schlieren images in Fig. 15 clearly illustrate the phenomenon of vortex splitting. It is evident that even the coarsest mesh can effectively capture the complex flow structures. As the resolution is further increased, finer features (e.g., Kelvin–Helmholtz instabilities near the slip line connected to the vortex) become more pronounced. The density profiles along $y = 0.4001$ (passing through the vortex center) are shown in Fig. 16, along with reference data[59] computed by the finite volume method using 8100×3000 cells. The results show that the current scheme provides accurate results without visible numerical oscillations, and it reaches a good agreement with the reference data. In the computation using 2000×1000 cells, the solution closely overlaps with that of the computation using the finest mesh. This case highlights the ability of the upwind CESE scheme in simulating complex phenomena involving both discontinuous and smooth flow structures.

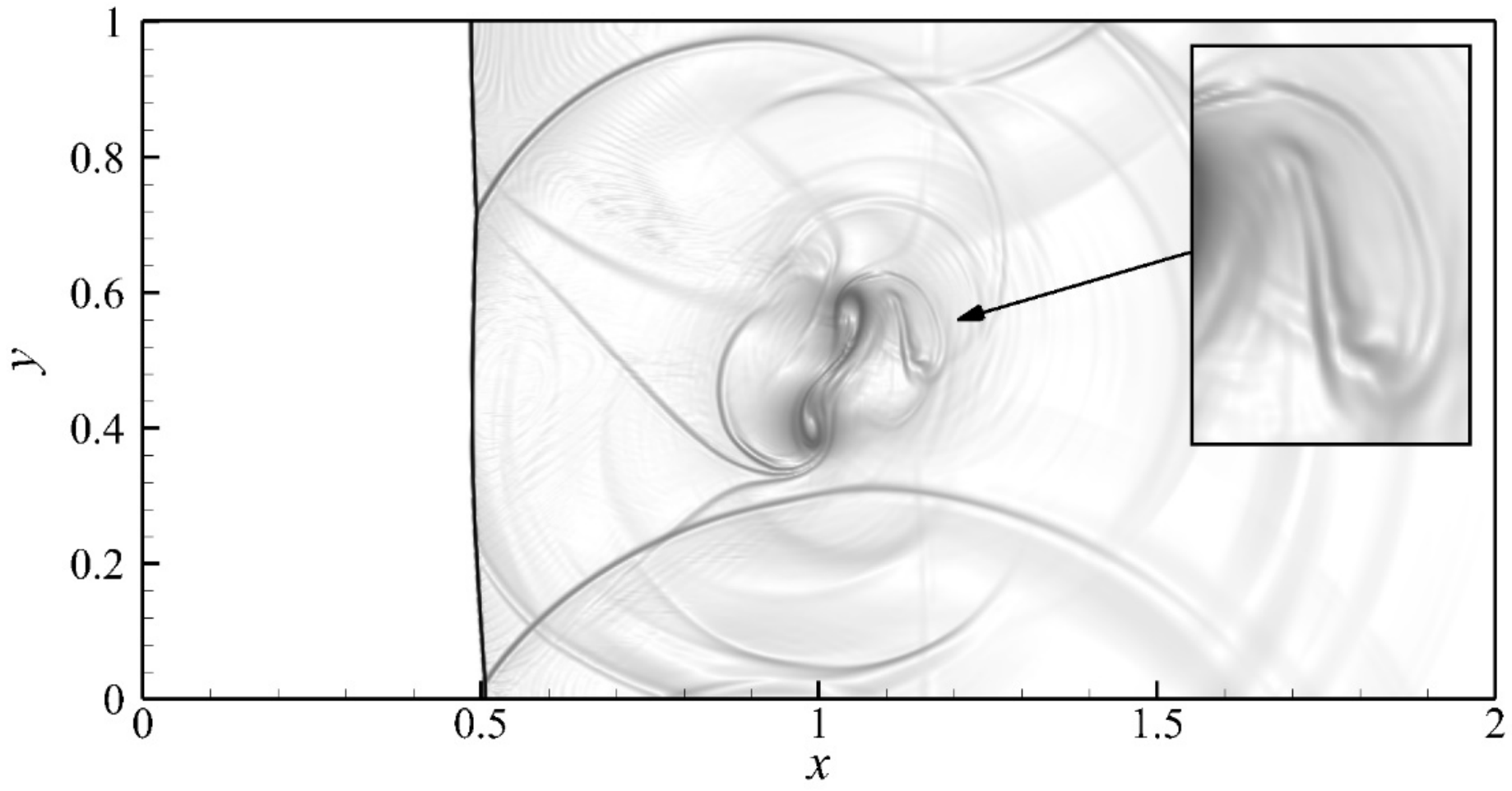

(a) 1000×500 cells

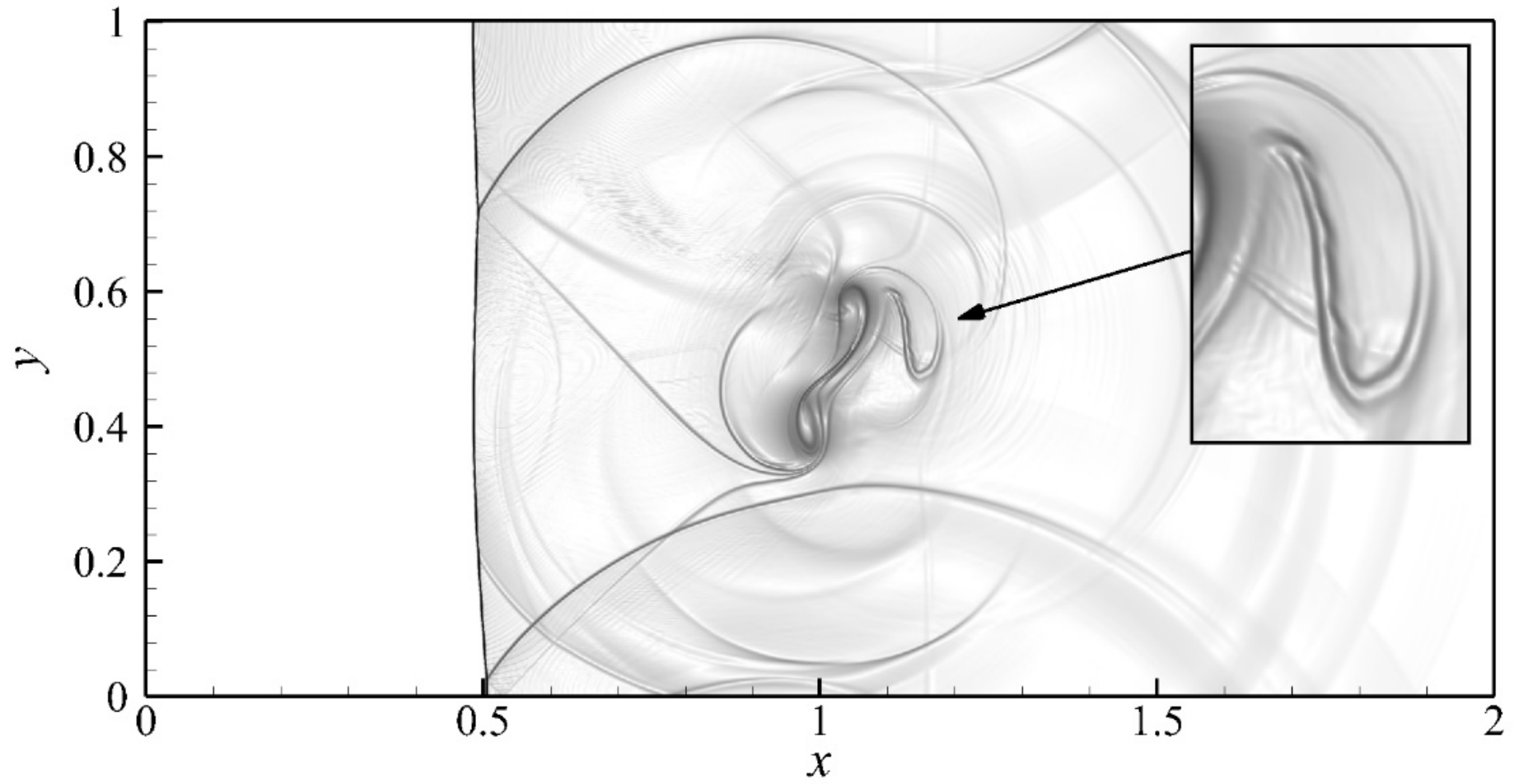

(b) 2000×1000 cells

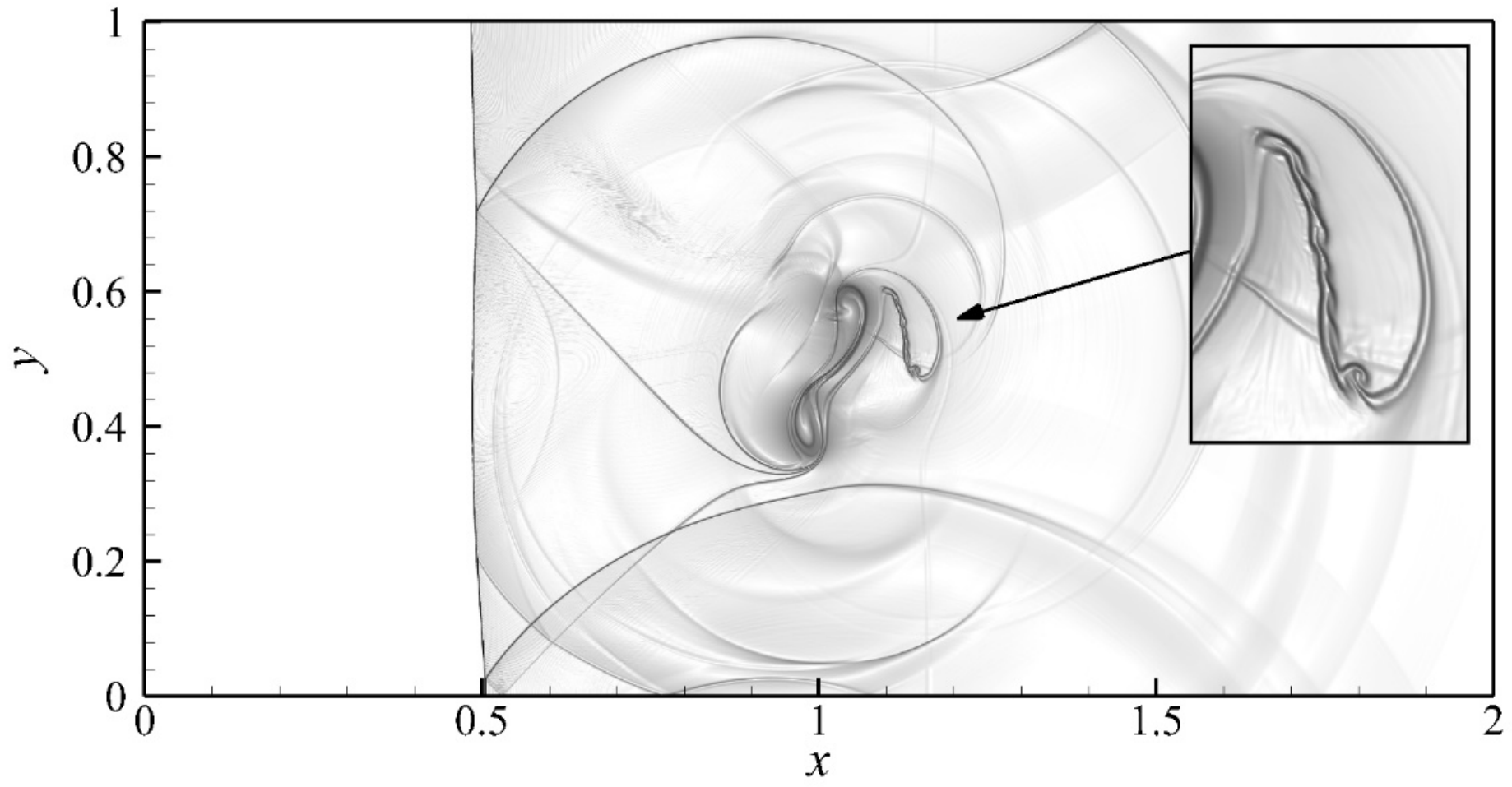

(c) 4000×2000 cells

FIG. 15 Numerical schlieren of the strong shock–vortex interaction problem at $t = 0.7$.

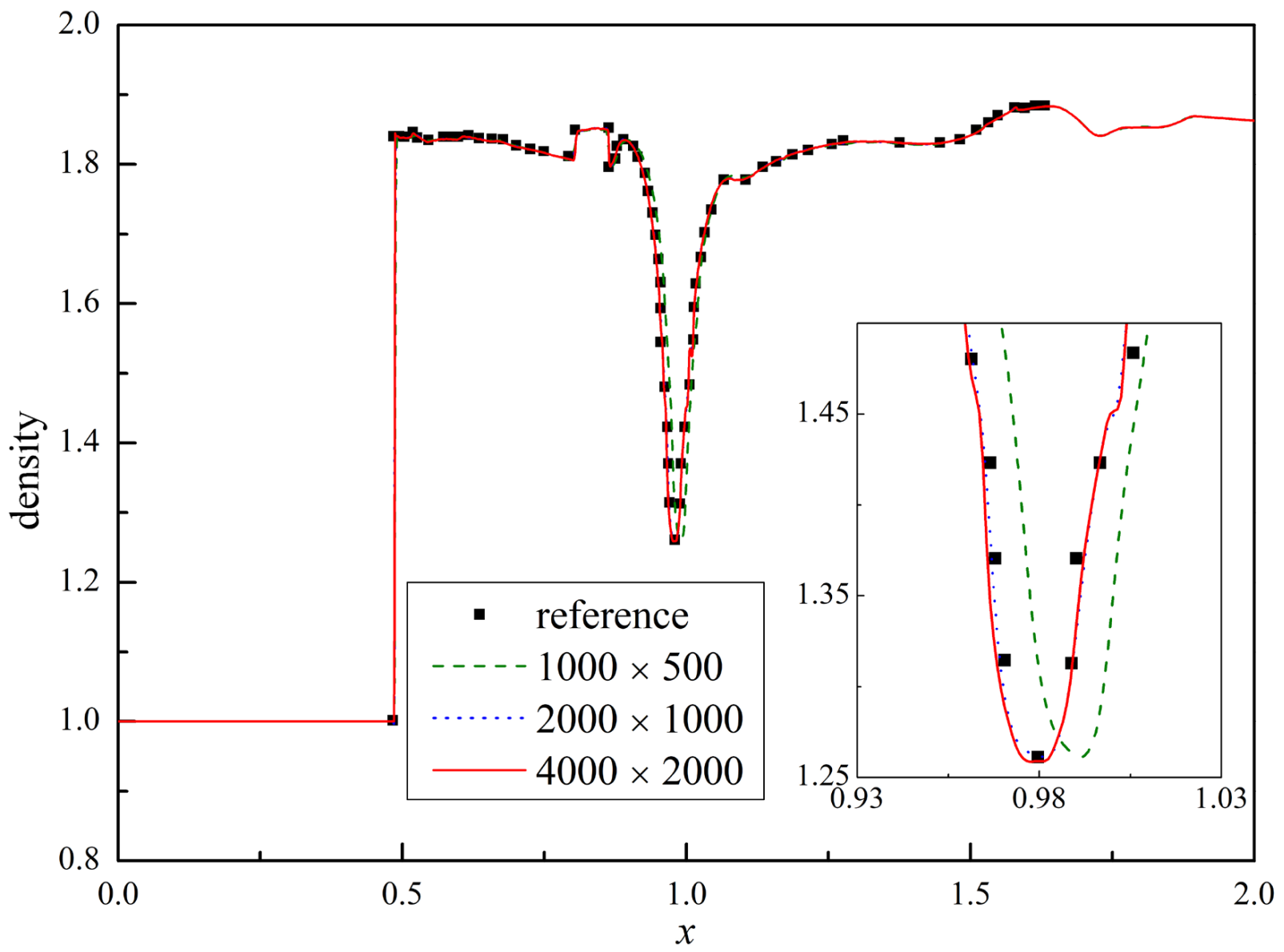

FIG. 16. Density profiles extracted at $y = 0.4001$ and $t = 0.7$ in the strong shock–vortex interaction problem. Lines: numerical results of the upwind CESE computations with different meshes. Symbols: reference data from Ref. 59.

## H. Shock–bubble interaction

The interaction of a moving planar shock wave and a cylindrical bubble is a fundamental case in the study of shock propagation and interfacial instability. Due to recent advancements in experimental techniques, various phenomena in shock–bubble interactions have been reported, as documented in high-quality experimental images and data. In this subsection, the upwind CESE scheme is used to numerically reproduce an experiment conducted by Ding et al.,[64] which investigated the interaction of an air shock and a cylindrical bubble filled with heavy gas. Comparisons between the numerical and experimental results can effectively assess the performance of the upwind CESE scheme in simulating compressible multi-fluid flows.

Based on the settings in the experiment, the problem is defined on a 2D domain [−50 mm, 150 mm] × [−70 mm, 70 mm]. The upper and lower boundaries represent the walls of the shock tube, which are assumed to be inviscid in the numerical simulation. The left and right sides of the domain are treated as non-reflecting boundaries. At the initial time, an incident shock moving from left to right at a Mach number of $M_s = 1.2$ is set up according to the Rankine–Hugoniot relations. The fluids in front of the shock wave are stationary, with an ambient pressure of 101325 Pa. A heavy gas bubble, with a diameter of 35 mm and center coordinates ($x = 0$, $y = 0$), is embedded in the pre-shock air. Before the passage of the shock, the densities of the air and the bubble content are $\rho_1 = 1.205$ kg/m$^3$ and $\rho_b = 4.859$ kg/m$^3$, respectively. Both fluids follow the perfect-gas equation of state with constant specific heat ratios of $\gamma_1 = 1.4$ and $\gamma_b = 1.117$, respectively. The physical models, governing equations, and the corresponding upwind CESE scheme have been described in Sec. II C and the references therein.

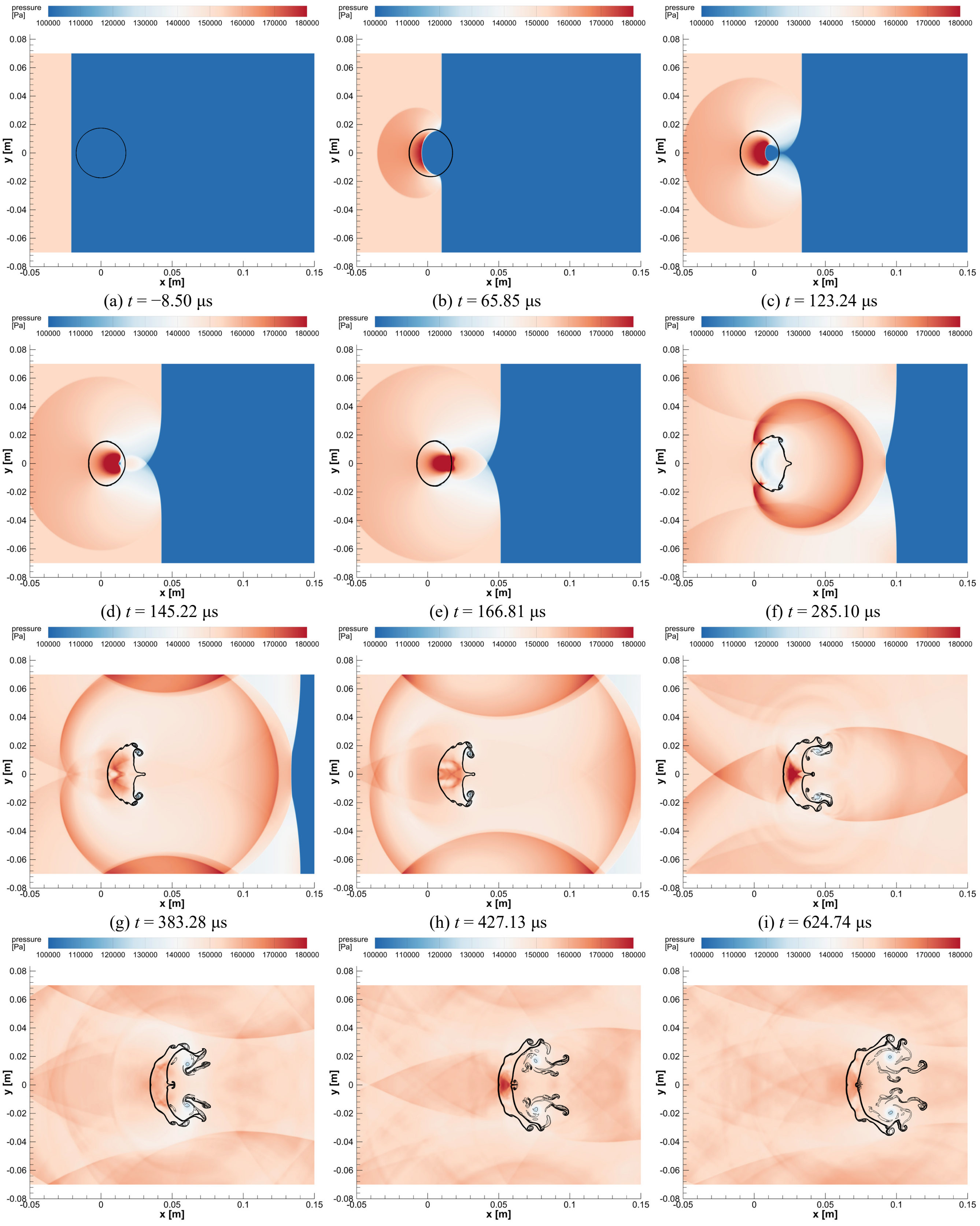

(a) $t = -8.50$ μs
(b) $t = 65.85$ μs
(c) $t = 123.24$ μs
(d) $t = 145.22$ μs
(e) $t = 166.81$ μs
(f) $t = 285.10$ μs
(g) $t = 383.28$ μs
(h) $t = 427.13$ μs
(i) $t = 624.74$ μs
(j) $t = 826.91$ μs
(k) $t = 1026.45$ μs
(l) $t = 1225.86$ μs

FIG. 17. Flow fields at different times in the shock–bubble interaction. Colors: pressure. Black lines: contour lines (nine equally spaced levels from 0.1 to 0.9) of the volume fraction of the first fluid (ambient air) in the two-fluid model.

The numerical simulation employs a uniform Cartesian mesh with a cell size of $\Delta x = \Delta y = 0.2$ mm, resulting in a total of 1000×700 cells. The time-dependent solution is advanced using a time step size determined by a CFL number of 0.3 and stops at $t = 1300$ μs. In Fig. 17, a series of snapshots showing the pressure field and the interface morphology are displayed. Note that the time $t$ is calibrated so that the definition of $t = 0$ is consistent with that of the experiment,[64] i.e. $t = 0$ is the very instant that the incident shock first contacts the bubble. As the system evolves, the reflection, refraction, diffraction, and focusing of shock waves are captured properly in the numerical results. The movement and deformation of the interface are also captured sharply. Fine structures, such as the penetrating jet at the rear of the bubble, a pair of main vortices, and numerous small-scale vortices, are resolved in Fig. 17. Good agreement is achieved between Fig. 17 and the experimental images in Ref. 64.

For the purpose of quantitatively assessing the CESE simulation, the width of the bubble (the span in the $y$-direction) is tracked as it evolves over time. The history of the bubble width is compared with the measurements based on the experimental data. As the incident shock wave passes through, the bubble shrinks because of the high post-shock pressure, causing the width to decrease before $t = 145$ μs. Subsequently, the evolution of the bubble is dominated by the formation and growth of vortices, leading to a gradual increase in the width. As shown in Fig. 18, the numerical results closely match the experimental data, highlighting the effectiveness of the upwind CESE scheme in simulating compressible multi-fluid flows.

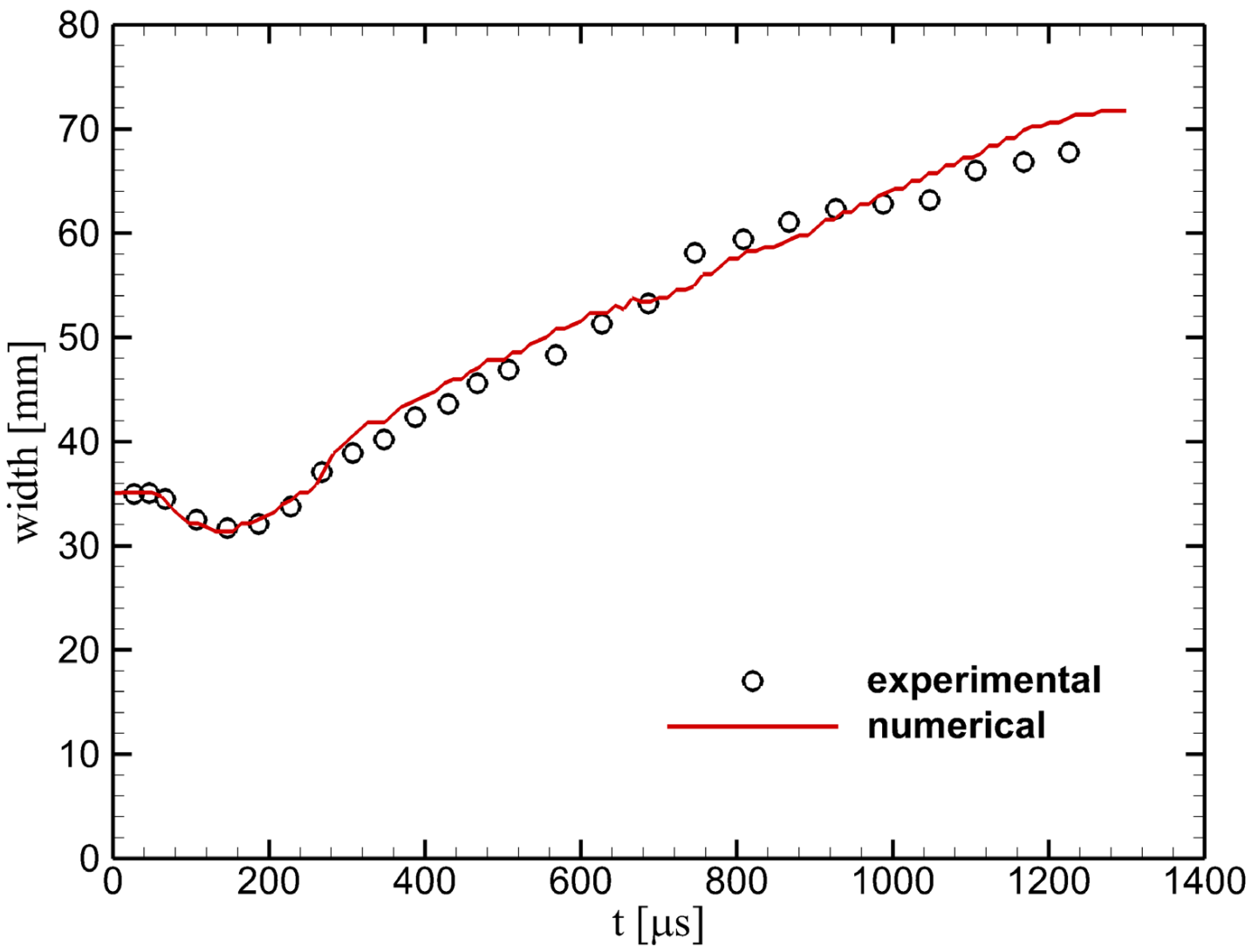

FIG. 18. Variation in the width of the heavy gas bubble during the shock–bubble interaction. The numerical results of the present study are compared to the experimental data in Ref. 64.

## V. CONCLUDING REMARKS

A relatively novel numerical framework for hyperbolic conservation laws, called the upwind CESE scheme, is analyzed to elucidate its numerical properties. The underlying modified equation of the upwind CESE scheme for a one-dimensional linear scalar advection equation ($u_t + au_x = 0$) is derived using Taylor expansions and an elimination procedure. The leading error term $(a\Delta x^2/48)(1- v^2)u_{xxx}$ verifies the second-order accuracy of the scheme, and this error possesses a smaller coefficient compared to the Lax–Wendroff scheme. Additionally, the spectral properties of the upwind CESE scheme are revealed by analyzing the propagation of all Fourier modes supported on the computational mesh. As the number of time steps approaches infinity, the dispersion and dissipation errors of the scheme in wavenumber space can be conveniently presented through the modified-wavenumber analysis. The modified wavenumber $\Phi$, and thus the dispersion and dissipation characteristics of the upwind CESE scheme, are expressed as functions of the CFL number $v = a\Delta t/\Delta x$ and the reduced wavenumber $\varphi = k\Delta x$. The upwind CESE scheme demonstrates satisfactory resolution for a wide range of wavenumbers.

To support the findings of the analysis, benchmark tests and applications to typical compressible flows are performed. The results obtained using the upwind CESE scheme agree well with reference solutions, including exact solutions, numerical results obtained using other methods, and experimental data. The upwind CESE scheme captures discontinuities sharply with good robustness, while preserving small-scale flow features with high resolution.

## ACKNOWLEDGMENTS

This research is supported by the National Natural Science Foundation of China (Grant No. 12302388), the Opening Project of the State Key Laboratory of Explosion Science and Technology (Beijing Institute of Technology, KFJJ23-20M), and the Hong Kong Research Grants Council (Grant No. 15217622).

## AUTHOR DECLARATIONS

### Conflict of Interest

The authors have no conflicts to disclose.

### Author Contributions

**Yazhong Jiang:** Conceptualization (lead); Data curation (equal); Formal analysis (lead); Funding acquisition (equal); Investigation (lead); Methodology (lead); Project administration (equal); Resources (equal); Software (equal); Supervision (equal); Validation (equal); Visualization (equal); Writing–original draft (equal); Writing–review & editing (equal). **Lisong Shi:** Data curation (equal); Funding acquisition (equal); Project administration (equal); Resources (equal); Software (equal); Supervision (equal); Validation (equal); Visualization (equal); Writing–original draft (equal); Writing–review & editing (equal). **Chih-Yung Wen:** Funding acquisition (equal); Project administration (equal); Resources (equal); Supervision (equal); Writing–review & editing (equal).

## DATA AVAILABILITY

The data that support the findings of this study are available from the corresponding author upon reasonable request.

## APPENDIX: OPERATIONS TO DERIVE THE MODIFIED EQUATION

From Eq. (47), it can be inferred that $g(x, t)$ and its derivatives are $O(\Delta t, \Delta x)$. Moreover, the following expressions can be obtained:

$$\Delta x^3 g_{xxx} = O(\Delta t^4, \Delta x^4)\,, \tag{A1}$$

$$\Delta x^2 g_{xx} = \frac{1}{4}\Delta x^3 u_{xxx} + O(\Delta t^4, \Delta x^4)\,, \tag{A2}$$

$$\Delta t \Delta x g_{tx} = \frac{1}{4}\Delta t \Delta x^2 u_{txx} + O(\Delta t^4, \Delta x^4)\,, \tag{A3}$$

$$\Delta x g_x = \frac{1}{4}\Delta x^2 u_{xx} + \frac{(\nu^2+\nu+4)}{48\nu}\Delta x^3 u_{xxx} - \frac{2\nu^2+2\nu-1}{12\nu(1-\nu)}\Delta t \Delta x g_{tx} + \frac{(\nu-2)(2\nu^2-\nu+2)}{12\nu(1-\nu)}\Delta x^2 g_{xx} + O(\Delta t^4, \Delta x^4)\,. \tag{A4}$$

Substituting Eqs. (A2) and (A3) into Eq. (A4) leads to

$$\Delta x g_x = \frac{1}{4}\Delta x^2 u_{xx} + \frac{\nu^2-5\nu+1}{48(1-\nu)}\Delta x^3 u_{xxx} - \frac{2\nu^2+2\nu-1}{48\nu(1-\nu)}\Delta t \Delta x^2 u_{txx} + O(\Delta t^4, \Delta x^4)\,. \tag{A5}$$

Using Eqs. (A1), (A2), and (A5), $g_x$, $g_{xx}$, and $g_{xxx}$ in Eq. (46) can be eliminated. Thus, Eq. (46) can be simplified as

$$\Delta t u_t + \nu \Delta x u_x = \frac{(\nu+1)(2\nu^4-5\nu^3+6\nu^2+10\nu-4)}{144\nu}\Delta x^3 u_{xxx} + \frac{(\nu+1)(\nu^2-2\nu+4)(2\nu^2+2\nu-1)}{144\nu^2}\Delta t \Delta x^2 u_{txx} + O(\Delta t^4, \Delta x^4)\,. \tag{A6}$$

To eliminate the temporal derivative $u_{txx}$ in Eq. (A6), Eq. (A6) itself is used to show that

$$\Delta t \Delta x^2 u_{txx} = -\nu \Delta x^3 u_{xxx} + O(\Delta t^5, \Delta x^5)\,. \tag{A7}$$

Hence, Eq. (A6) can be further simplified as

$$\Delta t u_t + \nu \Delta x u_x = \frac{\nu(1+\nu)(1-\nu)}{48}\Delta x^3 u_{xxx} + O(\Delta t^4, \Delta x^4)\,. \tag{A8}$$

Finally, by dividing both sides of Eq. (A8) by $\Delta t$ and using $\nu = a\Delta t/\Delta x$, the resulting equation is Eq. (48) in Sec. III B.

## REFERENCES

[1] S.-C. Chang and W.-M. To, "A new numerical framework for solving conservation laws—the method of space–time conservation element and solution element," NASA TM-104495 (1991).

[2] S.-C. Chang, "The method of space–time conservation element and solution element—A new approach for solving the Navier–Stokes and Euler equations," J. Comput. Phys. **119**, 295–324 (1995).

[3] Z. Zhai, W. Li, T. Si, X. Luo, J. Yang, and X. Lu, "Refraction of cylindrical converging shock wave at an air/helium gaseous interface," Phys. Fluids **29**, 016102 (2017).

[4] K. B. M. Q. Zaman, M. D. Dahl, T. J. Bencic, and C. Y. Loh, "Investigation of a 'transonic resonance' with convergent-divergent nozzles," J. Fluid Mech. **463**, 313–343 (2002).

[5] Z. Zhang, C.-Y. Wen, Y. Liu, D.-L. Zhang, and Z. Jiang, "Application of CE/SE method to gas-particle two-phase detonations under an Eulerian–Lagrangian framework," J. Comput. Phys. **394**, 18–40 (2019).

[6] Y. Liang, Z. Zhai, X. Luo, and C.-Y. Wen, "Interfacial instability at a heavy/light interface induced by rarefaction waves," J. Fluid Mech. **885**, A42–A60 (2020).

[7] H. Shen, C.-Y. Wen, M. Parsani, and C.-W. Shu, "Maximum-principle-satisfying space–time conservation element and solution element scheme applied to compressible multifluids," J. Comput. Phys. **330**, 668–692 (2017).

[8] C.-Y. Wen, H. Saldívar Massimi, H. Shen, "Extension of CE/SE method to non-equilibrium dissociating flows," J. Comput. Phys. **356**, 668–692 (2018).

[9] X. Feng, C. Xiang, D. Zhong, Y. Zhou, L. Yang, and X. Ma, "SIP-CESE MHD model of solar wind with adaptive mesh refinement of hexahedral meshes," Comput. Phys. Commun. **185**, 1965–1980 (2014).

[10] Y. Jiang, C.-Y. Wen, and D.-L. Zhang, "Space–time conservation element and solution element method and its applications," AIAA J. **58**, 5408–5430 (2020).

[11] C.-Y. Wen, Y. Jiang, and L. Shi, *Space–Time Conservation Element and Solution Element Method* (Springer, 2023).

[12] S.-C. Chang, X.-Y. Wang, and C.-Y. Chow, "The space–time conservation element and solution element method: A new high-resolution and genuinely multidimensional paradigm for solving conservation laws," J. Comput. Phys. **156**, 89–136 (1999).

[13] S.-C. Chang, "Courant number insensitive CE/SE schemes," in *38th AIAA/ASME/SAE/ASEE Joint Propulsion Conference & Exhibit*, Indiana, USA, (2002).

[14] S.-C. Chang, Y. Wu, V. Yang, and X.-Y. Wang, "Local time-stepping procedures for the space–time conservation element and solution element method," Int. J. Comput. Fluid Dyn. **19**(5), 359–380 (2005).

[15] G. Wang, D. Zhang, K. Liu, and J. Wang, "An improved CE/SE scheme for numerical simulation of gaseous and two-phase detonations," Comput. Fluids **39**, 168–177 (2010).

[16] H. Shen, R. A. Al Jahdali, and M. Parsani, "A class of high-order weighted compact central schemes for solving hyperbolic conservation laws," J. Comput. Phys. **466**, 111370 (2022).

[17] C.-L. Chang, "Time-accurate, unstructured-mesh Navier–Stokes computations with the space–time CESE method," in *42nd AIAA/ASME/SAE/ASEE Joint Propulsion Conference & Exhibit*, California, USA, (2006).

[18] C. Jiang, S. Cui, and X. Feng, "Solving the Euler and Navier–Stokes equations by the AMR–CESE method," Comput. Fluids **54**, 105–117 (2012).

[19] S.-C. Chang, C.-L. Chang, and J. C. Yen, "Recent developments in the CESE method for the solution of the Navier–Stokes equations using unstructured triangular or tetrahedral meshes with high aspect ratio," in *21st AIAA Computational Fluid Dynamics Conference*, California, USA, (2013).

[20] H. Shen and M. Parsani, "Positivity-preserving CE/SE schemes for solving the compressible Euler and Navier–Stokes equations on hybrid unstructured meshes," Comput. Phys. Commun. **232**, 165–176 (2018).

[21] H. Shen and M. Parsani, "A rezoning-free CESE scheme for solving the compressible Euler equations on moving unstructured meshes," J. Comput. Phys. **397**, 108858 (2019).

[22] L. Shi, C. Zhang, and C.-Y. Wen, "Adaptive mesh refinement algorithm for CESE schemes on quadrilateral meshes," arXiv:2409.01562v1.

[23] S.-C. Chang, "A new approach for constructing highly stable high order CESE schemes," in *48th AIAA Aerospace Sciences Meeting Including the New Horizons Forum and Aerospace Exposition*, Florida, USA, (2010).

[24] D. L. Bilyeu, S.-T. J. Yu, Y.-Y. Chen, and J.-L. Cambier, "A two-dimensional fourth-order unstructured-meshed Euler solver based on the CESE method," J. Comput. Phys. **257**, 981–999 (2014).

[25] H. Shen, C.-Y. Wen, K. Liu, and D.-L. Zhang, "Robust high-order space–time conservative schemes for solving conservation laws on hybrid meshes," J. Comput. Phys. **281**, 375–402 (2015).

[26] Y. Yang, X.-S. Feng, and C.-W. Jiang, "A high-order CESE scheme with a new divergence-free method for MHD numerical simulation," J. Comput. Phys. **349**, 561–581 (2017).

[27] H. Shen, C.-Y. Wen, and D.-L. Zhang, "A characteristic space–time conservation element and solution element method for conservation laws," J. Comput. Phys. **288**, 101–118 (2015).

[28] H. Shen and C.-Y. Wen, "A characteristic space–time conservation element and solution element method for conservation laws. II Multidimensional extension," J. Comput. Phys. **305**, 775–792 (2016).

[29] A. Rehman, I. Ali, and S. Qamar, "An upwind space–time conservation element and solution element method for solving dusty gas flow model," Results Phys. **7**, 3678–3686 (2017).

[30] Y. Yang, X.-S. Feng, and C.-W. Jiang, "An upwind CESE scheme for 2D and 3D MHD numerical simulation in general curvilinear coordinates," J. Comput. Phys. **371**, 850–869 (2018).

[31] H. Shen and M. Parsani, "The role of multidimensional instabilities in direct initiation of gaseous detonations in free space," J. Fluid Mech. **813**, R4 (2017).

[32] B. Guan, Y. Liu, C.-Y. Wen, and H. Shen, "Numerical study on liquid droplet internal flow under shock impact," AIAA J. **56**(9), 3382–3387 (2018).

[33] X. Feng, *Magnetohydrodynamic Modeling of the Solar Corona and Heliosphere*. (Springer, 2020).

[34] W. Li, Y. X. Ren, G. Lei, and H. Luo, "The multi-dimensional limiters for solving hyperbolic conservation laws on unstructured grids," J. Comput. Phys. **230**, 7775–7795 (2011).

[35] E. F. Toro, *Riemann solvers and numerical methods for fluid dynamics: A practical introduction*. (Springer, 2009).

[36] K. Kitamura, *Advancement of shock capturing computational fluid dynamics methods: Numerical flux functions in finite volume method*. (Springer, 2020).

[37] E. F. Toro, "The HLLC Riemann solver," Shock Waves **29**, 1065–1082 (2019).

[38] D.W. Levy, K.G. Powell, and B. van Leer, "Use of a rotated Riemann solver for the two-dimensional Euler equations," J. Comput. Phys. **106**, 201–214 (1993).

[39] Y. X. Ren, "A robust shock-capturing scheme based on rotated Riemann solvers," Comput. Fluids **32**, 1379–1403 (2003).

[40] H. Nishikawa and K. Kitamura, "Very simple, carbuncle-free, boundary-layer-resolving, rotated-hybrid Riemann solvers," J. Comput. Phys. **227**, 2560–2581 (2008).

[41] K.-M. Shyue, "An efficient shock-capturing algorithm for compressible multicomponent problems," J. Comput. Phys. **142**, 208–242 (1998).

[42] R. Saurel and R. Abgrall, "A simple method for compressible multifluid flows," SIAM J. Sci. Comput. **21**, 1115–1145 (1999).

[43] R. F. Warming and B. J. Hyett, "The modified equation approach to the stability and accuracy analysis of finite-difference methods," J. Comput. Phys. **14**, 159–179 (1974).

[44] S. K. Lele, "Compact finite difference schemes with spectral-like resolution," J. Comput. Phys. **103**, 16–42 (1992).

[45] C. K. W. Tam and J. C. Webb, "Dispersion-relation-preserving finite difference schemes for computational acoustics," J. Comput. Phys. **107**, 262–281 (1993).

[46] S. Pirozzoli, "On the spectral properties of shock-capturing schemes," J. Comput. Phys. **219**, 489–497 (2006).

[47] X. Y. Hu, V. K. Tritschler, S. Pirozzoli, and N. A. Adams, "Dispersion-dissipation condition for finite difference schemes," arXiv:1204.5088v2.

[48] F. Zhao, X. Ji, W. Shyy, and K. Xu, "An acoustic and shock wave capturing compact high-order gas-kinetic scheme with spectral-like resolution," Int. J. Comput. Fluid Dyn. **34**, 731–756 (2020).

[49] B. Einfeldt, C. D. Munz, P. L. Roe, and B. Sjögreen, "On Godunov-type methods near low densities," J. Comput. Phys. **92**, 273–295 (1991).

[50] P. D. Lax and X.-D. Liu, "Solution of two-dimensional Riemann problems of gas dynamics by positive schemes," SIAM J. Sci. Comput. **19** (2), 319–340 (1998).

[51] P. Woodward and P. Colella, "The numerical simulation of two-dimensional fluid flow with strong shocks," J. Comput. Phys. **54**, 115–173 (1984).

[52] L. I. Sedov, *Similarity and Dimensional Analysis in Mechanics*. (Academic Press, New York, 1959).

[53] V. Daru and C. Tenaud, "Evaluation of TVD high resolution schemes for unsteady viscous shocked flows," Comput. Fluids **30** (1), 89–113 (2000).

[54] G. A. Sod, "A survey of several finite difference methods for systems of non-linear hyperbolic conservation laws," J. Comput. Phys. **27**, 1–31 (1978).

[55] G.-S. Jiang and C.-W. Shu, "Efficient implementation of weighted ENO schemes," J. Comput. Phys. **126**, 202–228 (1996).

[56] C.-W. Shu and S. Osher, "Efficient implementation of essentially non-oscillatory shock-capturing schemes, II," J. Comput. Phys. **83**, 32–78 (1989).

[57] V. Springel, "*E pur si muove*: Galilean-invariant cosmological hydrodynamical simulations on a moving mesh," Mon. Not. R. Astron. Soc. **401**(2), 791–851 (2010).

[58] C. Hu and C.-W. Shu, "Weighted essentially non-oscillatory schemes on triangular meshes," J. Comput. Phys. **150**, 97–127 (1999).

[59] M. Galbraith, S. Murman, C. Kim, P. Persson, K. Fidkowski, R. Glasby, et al. in 5th international workshop on high-order CFD methods AIAA science and technology forum and exposition, (2017), http://how5.cenaero.be.

[60] J. L. Ellzey, M. R. Henneke, J. M. Picone, and E. S. Oran, "The interaction of a shock with a vortex: Shock distortion and the production of acoustic waves," Phys. Fluids **7**(1), 172–184 (1995).

[61] A. Rault, G. Chiavassa, and R. Donat, "Shock-vortex interactions at high Mach numbers," J. Sci. Comput. **19**, 347–371 (2003).

[62] J. Cheng, Z. Du, X. Lei, Y. Wang, and J. Li, "A two-stage fourth-order discontinuous Galerkin method based on the GRP solver for the compressible Euler equations," Comput. Fluids **181**, 248–258 (2019).

[63] B. Liang, M. Li, X. Yang, X. Tang, and J. Ding, "A cell-centered spatiotemporal coupled method for the compressible Euler equations," Phys. Fluids **35**, 066110 (2023).

[64] J. Ding, Y. Liang, M. Chen, Z. Zhai, T. Si, and X. Luo, "Interaction of planar shock wave with three-dimensional heavy cylindrical bubble," Phys. Fluids **30**, 106109 (2018).